\documentclass[%
 aip,
jcp,
 amsmath,amssymb,
 groupedaddress,
 reprint,%
longbibliography,
]{revtex4-1}

\usepackage{graphicx}% Include figure files
\usepackage{dcolumn}% Align table columns on decimal point
\usepackage{bm}% bold math

\usepackage[utf8]{inputenc}
\usepackage{amsmath}
\usepackage[T1]{fontenc}
\usepackage{mathptmx}
\usepackage{etoolbox}
\usepackage{braket}
\usepackage{mycommands}
\usepackage{comment}
\usepackage{hyperref}
\usepackage{esint}
\usepackage{tabularx}

\newcommand{\Hs}{\hat{H}_\mathrm{S}}
\newcommand{\Hb}{\hat{H}_\mathrm{B}}
\newcommand{\Hsb}{\hat{H}_\mathrm{SB}}
\newcommand{\Hls}{\hat{H}_\mathrm{LS}}
\newcommand{\ii}{\mathrm{i}}
\newcommand{\dd}{\mathrm{d}}

\begin{document}

\preprint{AIP/123-QED}

\title{Open quantum--classical systems: 
A hybrid MASH master equation}
\author{Kasra Asnaashari}
\author{Jeremy O. Richardson}\email{jeremy.richardson@phys.chem.ethz.ch}
\affiliation{Institute of Molecular Physical Science, ETH Zurich, 8093 Zurich, Switzerland}

\date{\today}

\begin{abstract}
We propose a method which combines the quantum--classical mapping approach to surface hopping (MASH) with the dissipative quantum dynamics of the Lindblad master equation.
Like conventional surface-hopping methods, our approach is based on classical trajectories coupled to the dynamics of a quantum subsystem.
However, instead of evolving the subsystem wavefunction according to the time-dependent Schrödinger equation, we use stochastic quantum trajectories derived from secular Redfield theory. This enables the simulation of open quantum systems coupled simultaneously to Markovian quantum baths and anharmonic non-Markovian classical degrees of freedom. Applications to the spin--boson model and to the cavity-enhanced fluorescence of an electronically nonadiabatic molecule show excellent agreement with fully quantum-mechanical benchmarks.
\end{abstract}

\maketitle

\section{Introduction}\label{sec:intro}

    Nonadiabatic processes play a central role in photochemistry, materials science, molecular quantum optics and even certain biological processes, where quantum and classical degrees of freedom interact in the presence of complex environments.\cite{GonzalezLindh,quantum-biology} These environments can range from slow thermally fluctuating solvents that can often be treated classically,\cite{Warshel1976retinol} to fast quantum-mechanical environments, 
    such as high-frequency vibrations,\cite{Siders1981quantum,Siders1981inverted}
    electronic excitations at a metal surface,\cite{HeadGordon1995friction} spin baths,\cite{Redfield1965relaxation,spin-baths,nuclear-spin-bath} 
    and quantized radiation fields.\cite{quantum-optics,Carmichael1,Carmichael2} Accurately modeling the interplay of these effects within a single framework remains an open problem.

    Quantum--classical trajectory methods, such as surface hopping\cite{Tully1990hopping} and mapping approaches,\cite{Meyer1979nonadiabatic,Stock1997mapping,Liu2016mapping,Miller2016Faraday,spinmap,multispin,MASH} provide efficient and accurate descriptions of nonadiabatic dynamics in the presence of harmonic and anharmonic classical degrees of freedom but fail to correctly model quantum-mechanical environments, where they can lead to disastrous zero-point energy leakage.\cite{Habershon2009water}
    On the other hand, most master equations designed for open quantum systems, such as Redfield theory,\cite{Redfield1965relaxation} naturally describe dissipation and decoherence due to quantum baths, but they are typically limited to a perturbative and Markovian treatment. Non-perturbative methods such as the hierarchical equations of motion (HEOM)\cite{Tanimura2020HEOM} can in principle describe quantum systems interacting with non-Markovian environments, but in practice are limited to simple system--bath models within the harmonic approximation.
    
    Attempts have been made to include dissipative dynamics in quantum--classical methods by either introducing an extra hopping probability in stochastic surface-hopping methods,\cite{RSH}
    using a Lindblad master equation in surface hopping,\cite{LME-SH}
    adding a damping term to the population measurements,\cite{SL-FSSH} or explicitly including the quantum bath states in the system.\cite{IESH,IESH2,IESH3,FR-SH}
    There are also stochastic surface-hopping methods designed to simulate the dynamics of a classical master equation derived from a semiclassical treatment of a quantum master equation.\cite{CME,CME2,QCLE-CME}
    Similarly from the perspective of open quantum systems, there have been attempts to efficiently model slow nuclear modes using a static classical ensemble\cite{BerkelbachFrozen} or to combine HEOM with Ehrenfest dynamics.\cite{HEOMThoss,HEOMclassical} 

    We introduce a hybrid approach that combines a stochastic formulation of Redfield theory with the mapping approach to surface hopping (MASH).\cite{MASH,MASHreview} In this framework, classical degrees of freedom are treated explicitly via classical trajectories with deterministic hops induced by nonadiabatic coupling to the quantum subsystem, while quantum baths are treated implicitly through stochastic jumps. This results in a trajectory-based method that is computationally tractable for complex molecular systems embedded in a quantum environment.

\section{Theory}

In this section, we outline the theoretical framework for our hybrid method, which combines elements of Redfield theory with MASH\@. We begin with the standard description of open quantum systems, where a quantum subsystem interacts with an external environment.\cite{OpenQuantum} Under the Born--Markov approximation, the reduced system dynamics can be described by a Lindblad master equation, for which secular Redfield theory provides a microscopic derivation of the dissipative and coherent contributions\footnote{By coherent contributions, we mean the Lamb shift in Eq.~\eqref{eq:redfield_secular_me}} in terms of the correlation functions of the bath. This makes Redfield theory a powerful tool in capturing quantum effects of weakly-coupled high-frequency baths. We then discuss a useful reformulation of the master equation through stochastic unravelling. 

On the other hand, MASH is a mixed quantum--classical approach, which describes the interaction of the quantum subsystem to a large number of classical variables through a set of coupled deterministic equations of motion.\cite{MASH,MASHreview}
This explicit representation of the classical degrees of freedom makes MASH particularly useful in simulating nonadiabatic molecular dynamics, in which the anharmonic nuclear modes are treated classically while the electronic state is treated quantum mechanically. 

\begin{figure*}
        \centering
        \includegraphics[width=0.8\linewidth]{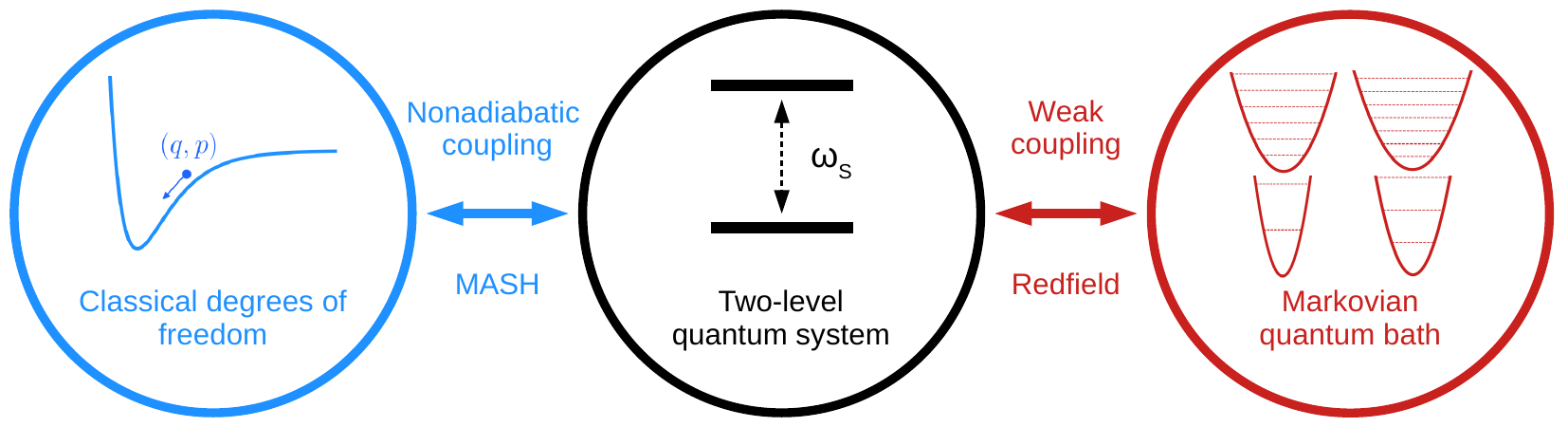}
        \caption{Schematic diagram of the hybrid Redfield--MASH method. The two-level quantum system is coupled simultaneously to a large number of classical degrees of freedom and a Markovian quantum bath. We use the MASH framework to treat the nonadiabatic coupling between the system and the classical degrees of freedom, and we use secular Redfield theory to treat the coupling to the quantum bath.}
        \label{fig:mash_redfield}
    \end{figure*}

Our goal is to merge these two approaches for problems with a quantum system coupled to both classical degrees of freedom and a quantum environment, as illustrated in Fig.~\ref{fig:mash_redfield}. In this way, each method can play to its strengths and together provide a unified description of nonadiabatic dynamics, dissipation and decoherence in open quantum--classical systems, which would be beyond the reach of either method alone. 

    \subsection{Secular Redfield Theory}\label{sec:redfield}

    As the dissipative quantum component of our hybrid approach, we consider a master equation within the framework of open quantum systems. 
    The Hamiltonian can be written as
    \begin{align}
        \hat{H} &= \Hs + \Hsb  + \Hb,
    \end{align}
    where $\Hs$ is the system Hamiltonian, $\Hb $ is the quantum bath Hamiltonian, and $\Hsb $ describes the system--bath coupling. In this work, we consider only two-level systems in a basis in which $\Hs = \omega_\mathrm{S}\hat{\sigma}_z / 2$ is diagonal. 
    Note that we employ units where $\hbar=1$ throughout.
    For simplicity, we assume a factorizable form of $\Hsb = \hat{A}\otimes\hat{B}$, where $\hat{A}$ acts on the system and $\hat{B}$ acts on the bath, although it would be trivial to extend all the theories presented in this paper to cases with non-factorizable system--bath couplings, $\Hsb=\sum_\alpha\hat{A}_\alpha\otimes\hat{B}_\alpha$.\cite{OpenQuantum}
    
    Redfield theory provides a microscopic derivation of the time-evolution of the system's reduced density matrix under the Born--Markov approximations.\cite{OpenQuantum} This assumes that the influence of the system on the reservoir is negligible (weak coupling) and that the bath excitations decay on timescales which are not resolved in the simulation (Markovian approximation). 
    
    In general, the Born--Markov approximations do not guarantee that the resulting master equation defines the generator of a dynamical semi-group.\cite{OpenQuantum}
    Therefore, a further secular approximation, which involves averaging over the rapidly oscillating terms in the master equation, is commonly applied to ensure a positive-definite density matrix. This approximation is valid when the timescale of the intrinsic evolution of the system is faster than the total relaxation time of the system coupled to the bath.
    
    Under these approximations, the effect of the bath is fully encoded in the Fourier transforms of the bath correlation functions
    \begin{align}\label{eq:bath_corr}
        \Gamma(\omega)&\equiv\int_0^\infty \dd t\, \mathrm{e}^{\ii \omega t} \, \Tr_\mathrm{B} \left[\hat{B}^\dagger(t) \hat{B}(0) \hat{\rho}_\mathrm{B}\right] ,
    \end{align}
    where ${\rm Tr}_\mathrm{B}[\cdot]$ is a trace over the bath degrees of freedom and $\hat{\rho}_\mathrm{B}$ is the bath's reduced density matrix. The time evolution of the system's reduced density matrix, $\hat{\rho}_\mathrm{S}$, takes the form of a Lindblad master equation: 
    \begin{align}
        \dot{\hat{\rho}}_\mathrm{S} = &-\ii \left[\Hs + \Hls, \hat{\rho}_\mathrm{S}\right] \nonumber \\
        &+ \sum_{\nu\in\{+,-,z\}} \gamma_\nu\big(\hat{\sigma}_\nu\hat{\rho}_\mathrm{S} \hat{\sigma}^\dagger_\nu - \tfrac{1}{2}\left[\hat{\sigma}^\dagger_\nu \hat{\sigma}_\nu, \hat{\rho}_\mathrm{S}\right]_+\big), \label{eq:redfield_secular_me}
    \end{align}
    where $[\cdot,\cdot]_+$ is the anti-commutator,
    \begin{align}
        \Hls  =& \sum_{\nu\in\{+,-,z\}}\xi_\nu  \hat{\sigma}^\dagger_\nu \hat{\sigma}_\nu
    \end{align}
    is the Lamb-shift Hamiltonian, which commutes with $\Hs$, and we define the coefficients 
    \begin{subequations}
    \begin{align}
        \gamma_\nu &= \left(\frac{{\rm tr}[\hat{\sigma}_\nu \hat{A}]}{{\rm tr}[\hat{\sigma}^\dagger_\nu\hat{\sigma}_\nu]}\right)^2 2\Re [\Gamma(\omega_\nu)], \label{eq:gamma_nu}\\
        \xi_\nu &= \left(\frac{{\rm tr}[\hat{\sigma}_\nu \hat{A}]}{{\rm tr}[\hat{\sigma}^\dagger_\nu\hat{\sigma}_\nu]}\right)^2 \Im [\Gamma(\omega_\nu)],
    \end{align}\label{eq:gamma_xi}
    \end{subequations}
    with $\omega_\pm = \mp\omega_\mathrm{S}$ and $\omega_z=0$.

    This formalism can be extended to treat time-dependent Hamiltonians. For this, we employ the real-valued adiabatic basis, $\ket{\Phi_0(t)}$ and $\ket{\Phi_1(t)}$, in which $\Hs(t)$ is always diagonal.  This introduces the time-derivative coupling $\tau(t)=\Braket{\Phi_1(t)|\frac{\dd}{\dd t}|\Phi_0(t)}$, leading to 
    \begin{align}
        \dot{\hat{\rho}}_\mathrm{S} = &-\ii \left[\Hs(t) + \Hls(t)  + \tau(t)\hat{\sigma}_y,\,\hat{\rho}_\mathrm{S}\right] \nonumber \\
        &+ \sum_{\nu\in\{+,-,z\}} \gamma_\nu(t)\big(\hat{\sigma}_\nu\hat{\rho}_\mathrm{S} \hat{\sigma}^\dagger_\nu - \tfrac{1}{2}\left[\hat{\sigma}^\dagger_\nu \hat{\sigma}_\nu, \hat{\rho}_\mathrm{S}\right]_+\big) ,\label{eq:redfield_secular_me_time}
    \end{align}
    where the coupling coefficients $\gamma_\nu(t)$ now depend on time, either directly due to $\hat{A}(t)$, or indirectly through the system frequency $\omega_S(t)$ and time-dependence of the adiabatic basis itself.
    The validity of this time-dependent master equation requires the system Hamiltonian to change slowly compared to the bath relaxation time, in addition to the other approximations used to derive secular Redfield theory (which must hold at all times).
    
    \subsection{Stochastic unravelling}\label{sec:stochastic}
    Stochastic unravelling of the master equation reformulates the evolution of the density matrix in terms of an ensemble of wavefunctions. Each wavefunction follows a stochastic ``quantum trajectory'' through the Hilbert space of the system and after averaging over many trajectories, we recover the original reduced density-matrix dynamics exactly.\cite{OpenQuantum}
    Although quantum trajectories are typically used to reduce the computational scaling from quadratic to linear in the size of the system Hilbert space, we will primarily use the master equation for two-state systems, where the scaling advantage of stochastic methods is not our concern.
    More importantly, the quantum-trajectory approach offers a picture that closely parallels trajectory-based classical and quantum--classical molecular dynamics. This will ultimately allow us to incorporate the Redfield treatment of quantum baths into a MASH framework, enabling a unified description of both nonadiabatic dynamics and dissipation.

    The stochastic unravelling of the master equation results in a piecewise deterministic process (PDP) in the Hilbert space of the system.
    The PDP associated with Eq.~\eqref{eq:redfield_secular_me_time} is\cite{OpenQuantum}
    \begin{align}
        \dd \psi(t) &= -\ii \mathcal{G}(\psi(t))\,\dd t \nonumber \\
        &\quad + \sum_{\nu}\left(\frac{\hat{\sigma}_\nu\psi(t)}{||\hat{\sigma}_\nu\psi(t)||} - \psi(t)\right)\dd N_\nu(t), \label{eq:stoch_dpsidt}
    \end{align}
    where the nonlinear deterministic evolution operator is
    \begin{align} \label{eq:G}
        \mathcal{G}(\psi(t)) &= \left(\Hs(t) + \Hls(t) \right)\psi(t)  - \frac{\ii }{2}\sum_\nu \gamma_\nu(t)\hat{\sigma}_\nu^\dagger \hat{\sigma}_\nu\psi(t) \nonumber \\
        &\quad+ \frac{\ii }{2} \sum_\nu \gamma_\nu(t) ||\hat{\sigma}_\nu\psi(t)||^2\psi(t),
    \end{align}
    and the second term of Eq.~\eqref{eq:stoch_dpsidt} represents the stochastic part of the evolution determined by the jump operators $\hat{\sigma}_\nu$ and the Poisson increments $\dd N_\nu(t)$, which satisfy
    \begin{subequations} \label{eq:stoch_jumps}
    \begin{align}
        \dd N_\mu(t) \, \dd N_\nu(t) &= \delta_{\mu\nu} \, \dd  N_\nu(t),
        \\
        \mathbb{E}[\dd  N_\nu(t)] &= \gamma_\nu(t) ||\hat{\sigma}_\nu\psi(t)||^2.
    \end{align}
    \end{subequations}
    In practice, one applies at most one of the jumps at each step with probability $\gamma_\nu(t)||\hat{\sigma}_\nu\psi(t)||^2 \dd t$. 

    In the case of a two-state system in the adiabatic representation, $\psi(t) = c_0(t)\ket{\Phi_0(t)} + c_1(t)\ket{\Phi_1(t)}$, the state can be represented by a spin vector on the Bloch sphere:
    \begin{subequations} \label{eq:spinvector}
    \begin{align}
        S_x &= 2{\rm Re}[c_1^*c_0], \\
        S_y &= 2{\rm Im}[c_1^*c_0], \\
        S_z &= |c_1|^2 - |c_0|^2.
    \end{align}
    \end{subequations}
    The nonlinear deterministic evolution [Eq.~\eqref{eq:G}] is equivalent to the following equations of motion for the spin vector:
    \begin{subequations}\begin{align}
        \dot{S}_x &= -\omega_{\rm LS}(t) S_y + 2\tau(t)S_z - \tfrac{1}{2}\left[\gamma_-(t) - \gamma_+(t)\right] S_xS_z, \\
        \dot{S}_y &= \omega_{\rm LS}(t) S_x - \tfrac{1}{2}\left[\gamma_-(t) - \gamma_+(t)\right] S_yS_z, \\
        \dot{S}_z &= -2\tau(t)S_x + \tfrac{1}{2}\left[\gamma_-(t) - \gamma_+(t)\right] (1 - S_z^2),
    \end{align} \label{eq:stoch_red_eom}\end{subequations}
    where $\omega_{\rm LS}(t)$ is the system frequency including the Lamb shift, i.e.\ the adiabatic energy gap of $\Hs(t) + \Hls(t) $. 
    As for the stochastic part of the evolution, at each step, we apply one of the jump operators $\hat{\sigma}_+, \hat{\sigma}_-, \hat{\sigma}_z$ with probability $\gamma_+|c_0|^2\dd t$, $\gamma_-|c_1|^2\dd t$, and  $\gamma_z\,\dd t$.  In the case of $\hat\sigma_\pm$, this resets the spin vector at one of the poles, and in the case of $\hat\sigma_z$, it rotates the spin vector by $\pi$ around the $z$-axis. 
    
    The reduced density matrix of the system can then be evaluated at any time from the ensemble of wavefunctions. 
    In the limit of a large number of quantum trajectories, this stochastic approach exactly reproduces the density-matrix dynamics of Eq.~\eqref{eq:redfield_secular_me_time} and is the inspiration for our hybrid Redfield--MASH method introduced in Sec.~\ref{sec:redmash}.
    
    \subsection{MASH}\label{sec:mash}
    In this section, we describe how to couple quantum and classical degrees of freedom in a consistent way using deterministic trajectories within the MASH formalism.\cite{MASH,MASHreview} This is useful both in simulating large molecular systems, where the nuclei can be treated classically and the electronic state quantum mechanically, or more generally, when simulating classical variables coupled to any two-level quantum system. 
    
    We begin with a quantum--classical Hamiltonian of the form 
    \begin{equation}
        \hat{H}(\bm{q},\bm{p}) = \sum_{j=1}^f \frac{p_j^2}{2} + \hat{V}(\bm{q}),
    \end{equation}
    where $\bm{q}$ and $\bm{p}$ are the mass-weighted coordinates and momenta for the $f$ classical degrees of freedom and $\hat{V}(\bm{q})$ is the potential operator (in the Hilbert space of the electronic states), which in the context of open quantum systems 
    includes $\Hs$, $\Hsb$ (which is assumed to be a function of $\bm{q}$ only) and the $\bm{q}$-dependent part of $\Hb$.
    
    For a two-level system, the electronic state is represented by a spin vector on the Bloch sphere [Eq.~\eqref{eq:spinvector}]. The nuclei are propagated on a single adiabatic potential energy surface according to Newton's equations of motion, while the spin vector evolves according to the time-dependent Schr\"odinger equation. This spin vector determines the adiabatic surface on which the classical coordinates propagate (called the active state). When the spin is in the upper/lower hemisphere, the upper/lower adiabat is used. This coupling between the nuclear motion and the electronic evolution allows MASH to capture nonadiabatic transitions in a deterministic manner.
    
    In the adiabatic representation, $\hat{V}(\bm{q}) = \bar{V}(\bm{q}) + \omega_{\rm S}(\bm{q}) \hat{\sigma}_z (\bm{q})/2$, the MASH equations of motion can be expressed as
    \begin{subequations}\begin{align}
        \dot{q}_j &= p_j \\
        \dot{p}_j &= -\pder{\bar{V}(\bm{q})}{q_j} - \frac{1}{2} \pder{\omega_\mathrm{S}(\bm{q})}{q_j}\sgn{S_z} \nonumber \\
        &\quad+ 2\omega_\mathrm{S} (\bm{q})d_j(\bm{q})S_x\delta(S_z) \label{eq:mash_force}\\
        \dot{S}_x &= -\omega_{\rm S}(\bm{q}) S_y + 2 \tau(\bm{q},\bm{p}) S_z \\
        \dot{S}_y &= \omega_{\rm S}(\bm{q}) S_x \\
        \dot{S}_z &= -2\tau(\bm{q},\bm{p}) S_x,\label{eq:mash_eom_sz}
    \end{align}\label{eq:mash_eom}\end{subequations}
    where $\tau(\bm{q},\bm{p}) = \bm{d}(\bm{q})\cdot\bm{p}$ and $\bm{d}(\bm{q}) = \Braket{\Phi_1(\bm{q}) | \nabla_{\bm{q}} | \Phi_0(\bm{q})}$ is the nonadiabatic coupling vector (NAC).

    The impulse in Eq.~\eqref{eq:mash_force} proportional to $\delta(S_z)$ is applied when the spin vector passes the equator. This results in a rescaling of the momentum when the trajectory has enough available energy to hop onto the other adiabat, and a reversal of the momentum when it does not. Both the rescaling and the reversal are applied along the direction of the NAC vector as described in previous work.\cite{MASH,MASHEOM}

    In addition to the equations of motion, it is necessary to specify how observables are mapped in the MASH framework. The Pauli spin operators in the locally adiabatic basis\footnote{Note the convention used for Pauli matrices as explained in Ref.~\onlinecite{MASHreview}} are mapped to spin vectors using the following procedure\cite{MASHreview}
    \begin{subequations}\label{MASHmapping}\begin{align}
        \hat{\sigma}_x(\bm{q})&\mapsto S_x, \\
        \hat{\sigma}_y(\bm{q})&\mapsto S_y, \\
        \hat{\sigma}_z(\bm{q})&\mapsto \sgn S_z.
    \end{align}\end{subequations}
	 In general, a quantum operator must be split up into its components before being mapped, e.g.\ $\hat{\mathcal{A}} = \sum_\mu \hat{\mathcal{A}}_\mu \mapsto \sum_\mu \mathcal{A}_\mu$.
    The MASH approximation to a quantum correlation function
    $C_{\mathcal{A}\mathcal{B}}(t)=\Tr[\hat{\rho}_0\hat{\mathcal{A}} \hat{\mathcal{B}}(t)]$
    is then
    \begin{align}
        C_{\mathcal{A}\mathcal{B}}^{\rm MASH}(t) &= \Braket{\rho_0(\bm{q},\bm{p}) \sum_{\mu\nu} \mathcal{A}_\mu(\bm{q},\bm{p},\bm{S})W_{\mu\nu}(t)\mathcal{B}_\nu(\bm{q}_t,\bm{p}_t,\bm{S}_t)} , \label{eq:mash_corr}
    \end{align}
    where $\rho_0(\bm{q},\bm{p})$ is the initial distribution in the classical phase space, and the subscript $t$ indicates a time-evolved variable. The ensemble average is
    \begin{align}
        \Braket{\cdots} &= \frac{1}{(2\pi)^f}\iiint \cdots\,\dd \bm{S}\,\dd \bm{q}\,\dd \bm{p},
    \end{align}
    where the integral over the Bloch sphere is defined as
    \begin{equation}
        \int \dd \bm{S}\cdots = \frac{1}{2\pi}\int_0^\pi\sin\theta\,\dd \theta\int_0^{2\pi} \dd \phi \cdots .
    \end{equation}
    Each trajectory is therefore initialized not only with a coordinate and momentum, but also with a spin vector sampled from the Bloch sphere. 
    Note that in this framework, individual trajectories with their associated spin vectors 
    do not represent physical states; it is only their ensemble average which has physical meaning through its prediction of correlation functions. 
    
    The weighting factor $W_{\mu\nu}$ in Eq.~\eqref{eq:mash_corr} depends on whether the operators $\mathcal{A}_\mu$ and $\mathcal{B}_\nu$ are population operators ${\rm P} = \{z,0,1,I\}$
    (i.e.\ proportional to $\hat{\sigma}_z$, electronic populations, the identity or pure nuclear operators)
    or coherence operators ${\rm C} = \{x, y\}$:
    \begin{align}
        W_{\mu\nu}(t) &= \begin{cases}
            3 & \mu\in {\rm C} \text{ and } \nu\in{\rm C}, \\
            3|S_z(t)| & \mu\in {\rm C} \text{ and } \nu\in{\rm P}, \\
            2 & \mu\in {\rm P} \text{ and } \nu\in{\rm C}, \\
            2|S_z(t)| & \mu\in {\rm P} \text{ and } \nu\in{\rm P}. \\
        \end{cases}\label{eq:mash_weights}
    \end{align}
    Note that these weighting factors 
    are different from those presented in the original MASH method of Ref.~\onlinecite{MASH}.
    One difference is that they are evaluated at time $t$, instead of at the start of the trajectory (an option already considered in previous work\cite{MASH,MASHrates}).
    In addition, the original MASH method used $W_{\rm CP} = 2$, where we propose $3|S_z(t)|$. 
    The reason that it is necessary for us to change the definition of the weighting factors is that,  
    as we show in Sec.~\ref{sec:redmash}, we employ non-unitary equations of motion for the spin vector in our hybrid method, whereas the original weighting factors were derived to capture the correct electronic dynamics under the assumption of unitary evolution.\cite{MASH,MASHreview}
    It has however been noted that the original definitions were not unique and that a number of different weighting factors and estimators can be derived and used to evaluate observables using MASH dynamics with minor differences in the results.\cite{mash-thesis}
    In Appendix~\ref{ap:weighting}, we demonstrate how the new weighting factors in Eq.~\eqref{eq:mash_weights} still reproduce the correct electronic dynamics given a predetermined nuclear path. In fact, these new weighting factors retain all of the desirable properties of the original MASH method (with one minor exception discussed in Appendix~\ref{ap:QCLE}), while allowing for non-unitary spin equations of motion.
    We thus expect the accuracy of the results with the new weighting functions to be similar to those of the original MASH method.
    A full numerical analysis is left for future work.

    MASH offers unique advantages that set it apart from previously proposed trajectory-based nonadiabatic dynamics. It exactly reproduces the quantum--classical Liouville equation (QCLE)\cite{Kapral2015QCL} up to first-order in time\cite{MASH,MASHreview} and relaxes to the correct thermal equilibrium distribution.\cite{thermalization} The MASH equations of motion are deterministic and time-reversible,\cite{MASHEOM} a property not shared by stochastic surface-hopping methods. Additionally, MASH yields accurate thermal rates for a wide range of models without the need for explicit decoherence corrections,\cite{MASHrates} and can accurately simulate electronic coherences.\cite{Mannouch2024coherence,MASHcoh} Like other surface-hopping algorithms, it can treat anharmonic problems and can be used to efficiently simulate \textit{ab initio} nonadiabatic molecular dynamics.\cite{Mannouch2024MASH,cyclobutanone}
    There have also been attempts to generalize MASH to systems with more than two quantum states,\cite{unSMASH,Runeson2023MASH,botticelli2025semi} although they break some of the good properties of the original method. 
    \cite{MASHreview}
	Nonetheless, they have been used successfully in certain situations.\cite{cyclobutanone,Runeson2024MASH,Runeson2024semiconductors,Runeson2025NQE}
    In this work, we focus only on the two-state version and leave possible multi-state generalizations for future studies.
    
    \subsection{Hybrid Redfield--MASH}\label{sec:redmash}
    
    Having introduced both the Redfield and MASH formalisms, we now combine them into a single framework that captures the strengths of both approaches.  Here, we consider a two-level electronic system not only weakly-coupled to a Markovian quantum environment but also coupled to classical nuclear degrees of freedom. 
    
    A natural first attempt might be to associate the spin vector of MASH with a density matrix, rather than a pure-state wavefunction, and evolve it using a master equation (as has been done in the context of FSSH\cite{LME-SH,QCLE-CME}). However, this approach will clearly fail to reproduce the correct thermal equilibrium distribution within the MASH mapping framework. This is because the thermalized density matrix will inevitably settle in the lower hemisphere of the Bloch sphere, such that all MASH trajectories would hop to the lower adiabatic state, regardless of the bath temperature. 
    
    To address this, we turn to the stochastic unravelling of the master equation, which like MASH is also a trajectory-based method, albeit a quantum trajectory. In Appendix~\ref{ap:proof}, we derive a PDP that reproduces the Redfield density-matrix evolution [Eq.~\eqref{eq:redfield_secular_me_time}] along a given nuclear path using an ensemble of spin vectors within the MASH framework. The deterministic part of the spin evolution is 
    \begin{subequations}\begin{align}
            \dot{S}_x &= -\omega_{\rm LS}(\bm{q}) S_y + 2\tau(\bm{q},\bm{p})S_z \nonumber \\
            &\quad- \tfrac{1}{2}\left[\gamma_-(\bm{q}) - \gamma_+(\bm{q})\right]S_x\sgn S_z, \\
            \dot{S}_y &= \omega_{\rm LS}(\bm{q}) S_x - \tfrac{1}{2}\left[\gamma_-(\bm{q}) - \gamma_+(\bm{q})\right]S_y \sgn S_z, \\
            \dot{S}_z &= -2\tau(\bm{q},\bm{p})S_x.\label{eq:spin_evol_sz}
    \end{align}\label{eq:spin_evol}\end{subequations}
    These equations closely parallel Eq.~\eqref{eq:stoch_red_eom}, except that $S_z$ is replaced by $\sgn S_z$ in the terms involving the bath effects [in the spirit of the MASH mapping procedure, Eq.~\eqref{MASHmapping}].
    Note that, here, the time-dependence of $\omega_\mathrm{S}$ and $\gamma_\pm$ appears through the classical coordinates, as the adiabatic energy gap and the adiabatic basis depend on $\bm{q}$. 
    In principle, this approach also allows for a position-dependent system--bath coupling operator, $\hat{A}(\bm{q})$, although we will not specifically consider this case in this work.

    Similar to the stochastic method described above, at each step of the algorithm, we randomly apply one of the jump operators $\hat{\sigma}_\pm$ or $\hat{\sigma}_z$ with the probability $\mathcal{P}_\pm(t)\dd t$ or $\mathcal{P}_z(t)\dd t$, given by
    \begin{subequations}\begin{align}
        \mathcal{P}_\pm(t) &= \gamma_\pm h(\mp S_z(t)), \\
        \mathcal{P}_z(t) &= \gamma_z.
    \end{align}\label{eq:hybrid_jumps}\end{subequations}
    The $\hat{\sigma}_\pm$ jump operators flip the active state. 
    However, unlike the standard stochastic Redfield, we do not restart the spin vector from the poles of the Bloch sphere but sample a new vector from the appropriate hemisphere according to the MASH framework.
    These resamplings also come with new weighting factors
    defined recursively after $n$ jumps as
    \begin{align}
        \hat{\sigma}_\pm &\Rightarrow\begin{cases}
            W_{\mu \mathrm{P}}^{(n)}(t) = W_{\mu \mathrm{P}}^{(n-1)}(t_n) 2 h(\pm S'_z)|S'_z(t)| \\
            W_{\mu \mathrm{C}}^{(n)}(t) = W_{\mu \mathrm{P}}^{(n-1)}(t_n) 2 h(\pm S'_z) \\
        \end{cases}, \label{eq:jump_weights}
    \end{align}
    where the $n$th jump occurs at time $t_n$, and $\bm{S}'$ is the newly sampled vector after the jump (which effectively must be sampled from the upper hemisphere for a $\hat\sigma_+$ jump and from the lower hemisphere for a $\hat\sigma_-$ jump). The derivation of these factors is presented in Appendix~\ref{ap:proof} based on the concept that a $\hat\sigma_\pm$ operator measures a population. At all times before the first $\hat{\sigma}_\pm$ jump, we use the weighting factors $W^{(0)}_{\mu\nu}(t)$ from Eq.~\eqref{eq:mash_weights}. 

    For $\hat{\sigma}_z$ jumps, we simply rotate the vector around the $z$-axis [$\bm{S}'=(-S_x,-S_y,S_z)$] and continue without reweighting the trajectory. 
    In Appendix~\ref{ap:proof_weight}, we discuss an alternative (and more computationally expensive) procedure, which involves resampling $\bm{S}'$ from the full Bloch sphere after $\hat{\sigma}_z$ jumps.  However, in Appendix~\ref{ap:proof_eom}, we show that the simpler approach of rotating the vector is sufficient to reproduce the correct reduced density-matrix dynamics.

    The nonlinear spin evolution of Eq.~\eqref{eq:spin_evol} does not conserve the norm of the vector and can bring the spin vector inside or outside the Bloch sphere.
    In fact, because of the condition $\gamma_+ = \exp[-\beta\omega_{\rm S}]\gamma_- < \gamma_-$, when the vector is pointing in the upper hemisphere, the norm of the vector grows, while it shrinks in the lower hemisphere.
    This does not cause a problem as the spin vectors in this method are not associated with physical wavefunctions. 
    Appendix~\ref{ap:proof} shows that we recover the correct observables on an ensemble level.
    
    Note that, the equation of motion for the stochastic Redfield approach [Eq.~\eqref{eq:stoch_red_eom}] includes an extra term in the $S_z$ dynamics compared to Eq.~\eqref{eq:spin_evol}. This term describes a continuous population decay to the lower state. 
    In our formulation, population transfer can happen only through stochastic jumps, which therefore have a higher probability of occurring at each step, according to Eq.~\eqref{eq:hybrid_jumps}, compared to their counterparts in Eq.~\eqref{eq:stoch_jumps}.
    This subtle difference is a consequence of putting the formalism into the MASH framework. However, both methods exactly reproduce the reduced density-matrix dynamics of secular Redfield theory [Eq.~\eqref{eq:redfield_secular_me_time}] despite the differences in the dynamics of individual trajectories.

    So far, we have only considered a reformulation of Redfield theory for an open quantum system.
    To construct the hybrid method, we now couple the system to classical degrees of freedom in addition to the quantum bath.  
    The classical modes are thus indirectly coupled to the quantum bath and, apart from the small Lamb shift, feel its effects only through the system dynamics.
    
    The dynamics of the classical variables $(\bm{q}, \bm{p})$ are defined naturally within the MASH framework. In particular, they evolve according to the forces of one of the adiabats, determined by the sign of $S_z$: 
    \begin{subequations}\begin{align}
        \dot{q}_j &= p_j, \label{eq:mash_qdot}\\
        \dot{p}_j &= -\pder{\bar{V}_{\rm LS}(\bm{q})}{q_j} -\frac{1}{2}\pder{\omega_{\rm LS}(\bm{q})}{q_j}\sgn{S_z} \nonumber \\
        &\quad+ 2\omega_\mathrm{LS} (\bm{q})d_j(\bm{q})S_x\delta(S_z), \label{eq:mash_force}
    \end{align}\end{subequations}
    where we additionally include the effect of the Lamb shift in the force.

\begin{figure}
        \centering
        \includegraphics[width=\linewidth]{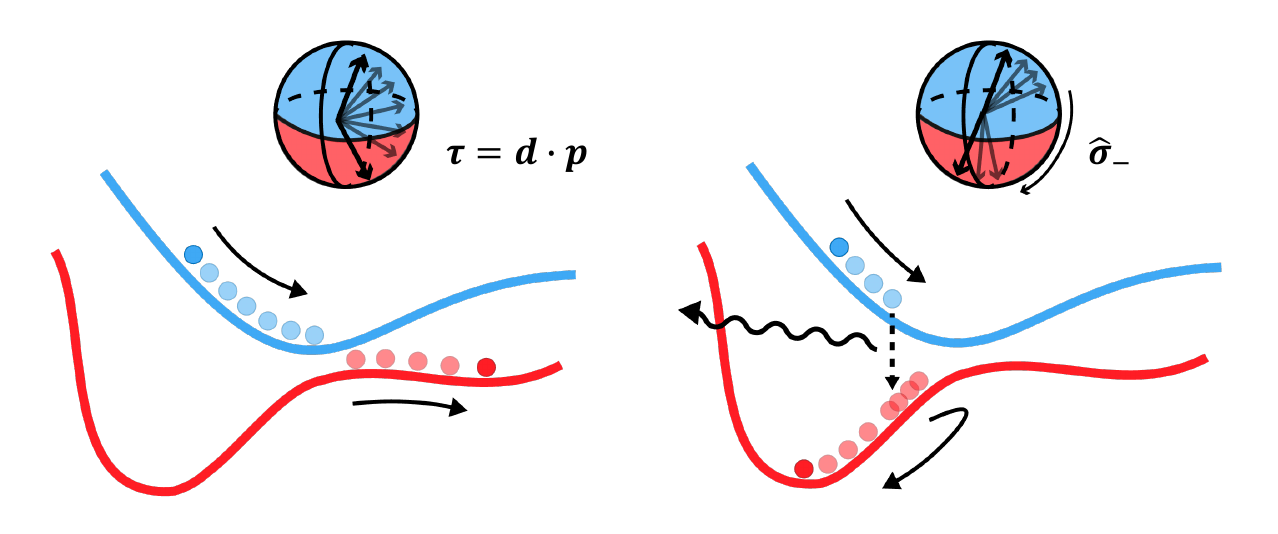}
        \caption{The two distinct mechanisms for changing active state in the hybrid method. Left: Energy-conserving nonadiabatic ``hops'' as the spin vector passes the equator, mediated by the coupling to the classical coordinates. Right: Stochastic ``jumps'' initiated by the Lindblad operators coupling the system to the quantum bath.}
        \label{fig:hops_v_jumps}
    \end{figure}

    Therefore, our new hybrid method has two distinct mechanisms of changing the active state, as illustrated in Fig.~\ref{fig:hops_v_jumps}. First, there are the deterministic ``hops'' caused by the NACs which couple the classical coordinates to the quantum subsystem.  Similarly to the original MASH method, these occur whenever $S_z$ crosses the equator of the Bloch sphere. 
    Second, there are the stochastic ``jumps'' that resample $\bm{S}$ in the other hemisphere. These resemble the Lindblad operators in a stochastically unraveled master equation. 
    
    As in the original MASH method, hopping between adiabatic surfaces is accompanied by a momentum rescaling along the direction of the NAC to conserve the available energy and a reflection in the case of frustrated hops.\cite{MASH} However, we do not rescale the momentum after jumps, because they are caused by the system--bath interactions and thus do not have to conserve the energy of the system. 

    In the same way that MASH is justified through its connection to the QCLE,\cite{MASH,MASHreview} the hybrid dynamics
    described in this section are consistent with the short-time limit of the partial Wigner transform of Eq.~\eqref{eq:redfield_secular_me_time}.
    Appendix \ref{ap:qcle-cme} explains the connection between the hybrid Redfield--MASH method and the secular-Redfield version of the ``QCLE inside a classical master equation (QCLE-CME)'' derived in Refs.~\onlinecite{QCLE-CME-friction,QCLE-CME} for full Redfield dynamics. 

\section{Results}\label{sec:results}
    In order to demonstrate that our hybrid method combines the advantages of MASH for non-Markovian classical degrees of freedom and the secular Redfield treatment of Markovian quantum baths, we apply it to two model systems for which benchmark results are available. First, we consider a spin--boson model with two baths, one that can be accurately modeled classically by MASH and one that can be accurately modeled with Redfield theory, while neither method alone can accurately capture the effects of both baths simultaneously. This assesses the ability of our hybrid method to treat both baths consistently and accurately.
    Second, we present an application to quantum optics, by simulating the cavity-enhanced fluorescence of a molecule (modeled by a single anharmonic vibrational mode) in competition with nonradiative transitions.

    To evolve the MASH trajectories, we use a second-order integrator proposed in Ref.~\onlinecite{MASHEOM} (in particular rev-pc-NACs). 
    The hybrid simulation uses a similar approach based on a symmetric splitting to alternate between propagating the classical degrees of freedom using the velocity-Verlet algorithm and the spin vector, which is either integrated deterministically or resampled according to the stochastic jump procedure. 
    Note that as long as $\sgn S_z$ does not change, the non-unitary deterministic spin evolution of Eq.~\eqref{eq:spin_evol} is linear, $\dot{\bm{S}} = \mathsf{\Omega}(\bm{q},\bm{p},\sgn S_z)\bm{S}$, such that the update for the spin vectors can be performed exactly as
    \begin{equation}
        \bm{S}(t+\Delta t)= \mathrm{e}^{\mathsf{\Omega}(\bm{q},\bm{p},\sgn S_z)\Delta t} \, \bm S(t).\label{eq:total_spin_evol}
    \end{equation}
    When hops occur, we use 10 bisections to find the exact hopping time (up to an error of $\Delta t/2^{10}$).\cite{Runeson2023MASH} 
    This allows us to use the exact spin evolution of Eq.~\eqref{eq:total_spin_evol} before and after the hopping time.
    Note that it may be possible to adapt specialized propagators for PDPs\cite{Kloeden1992} to our problem, which may further improve the efficiency of our hybrid method.

    Unlike the original MASH method, the weighting factors introduced in Eq.~\eqref{eq:mash_weights} and Eq.~\eqref{eq:jump_weights} are functions of $\bm{S}(t)$. 
    Assuming that trajectories are initialized on the upper adiabatic state (as is the case in Sec.~\ref{sec:spontaneous_emission}),
    a naive formula for the adiabatic populations would be
    \begin{align} \label{eq:naive}
        P_{0/1}^\mathrm{naive}(t) &= \frac{\langle \rho_0(\bm{q},\bm{p}) h(S_z) W_{\rm PP}(t) h(\mp S_z(t))\rangle}{\langle \rho_0(\bm{q},\bm{p}) h(S_z) W_{\rm PP}(0) \rangle}.
    \end{align}
    This leads to problems as they sum to a total population of
    \begin{equation}
        P_0^\mathrm{naive}(t) + P_1^\mathrm{naive}(t) = \frac{\langle \rho_0(\bm{q},\bm{p}) h(S_z) W_{\rm PP}(t) \rangle}{\langle \rho_0(\bm{q},\bm{p}) h(S_z) W_{\rm PP}(0) \rangle},
    \end{equation}
    which is not guaranteed to be equal to 1 for $t>0$.
    To rectify this issue, we use an alternative and more symmetric definition of the population operator, which has been introduced previously to fix a similar problem in other mapping approaches.\cite{identity}
    Expanding $\hat{\mathcal{B}}=\ket{\Phi_0}\bra{\Phi_0}=\half(\hat{I}-\hat\sigma_z)$ or $\hat{\mathcal{B}}=\ket{\Phi_1}\bra{\Phi_1}=\half(\hat{I}+\hat\sigma_z)$,
    and using the exact quantum-mechanical result for the trivial dynamics of the identity,
    we define the improved adiabatic populations as
    \begin{align}
        P_{0/1}(t) &= \half \left[1 \mp \frac{\langle \rho_0(\bm{q},\bm{p}) h(S_z) W_{\rm PP}(t) \sgn(S_z(t))\rangle}{\langle \rho_0(\bm{q},\bm{p}) h(S_z) W_{\rm PP}(0) \rangle} \right],\label{eq:sz_plus_1}
    \end{align}
    which can be easily shown to sum to 1. 
    
    Diabatic populations in Sec.~\ref{sec:spin-boson} are calculated in a similar way using sums and differences.  In particular, we define $\hat{\mathcal{A}}=\ketbra{a}{a}$ and $\hat{\mathcal{B}}=\ketbra{a}{a}-\ketbra{b}{b}$, such that the diabatic population of $a$ can be evaluated as
    \begin{align}
        P_a(t) = \half \left[1 + \frac{C^\mathrm{MASH}_\mathcal{AB}(t)}{C^\mathrm{MASH}_\mathcal{AB}(0)} \right] .\label{eq:diabatic_population}
    \end{align}
    The MASH correlation function 
    [Eq.~\eqref{eq:mash_corr}] is itself defined as a sum of nine terms (which can all be calculated from the same ensemble of trajectories) obtained by expanding both $\hat{\mathcal{A}}$ and $\hat{\mathcal{B}}$ in terms of position-dependent linear combinations of the adiabatic operators $\hat{I}$, $\hat{\sigma}_z$ and $\hat{\sigma}_x$.\footnote{Actually, only six of the nine terms are non-zero as $\hat{\mathcal{B}}$ has no identity component in this case.} 

    \subsection{Spin--boson models} \label{sec:spin-boson}
    We define a two-level system coupled to two bosonic baths, a slow classical (low-frequency) bath and a fast quantum (high-frequency) bath. The Hamiltonian in the diabatic basis, $\ket{a}$ and $\ket{b}$, is 
    \begin{subequations}\begin{align}
        \Hs &= \begin{pmatrix}
            \varepsilon & \Delta \\
            \Delta & -\varepsilon
        \end{pmatrix}, \\
        \hat{H}_{\mathrm{B}} &= \sum_{j,\alpha} \frac{\hat{p}^2_{j,\alpha}}{2} + \frac{1}{2}\sum_{j,\alpha}\omega_{j,\alpha}^2 \hat{q}^2_{j,\alpha}, \\
        \hat{H}_{\mathrm{SB}} &= \begin{pmatrix}
            1 & 0 \\
            0 & -1
        \end{pmatrix}\otimes\sum_{j,\alpha} c_{j,\alpha}\hat{q}_{j,\alpha},
    \end{align}\end{subequations}
    where $\alpha=\{\mathrm{c},\mathrm{q}\}$ for the classical and quantum baths, $\varepsilon$ is the energy bias, $\Delta$ is the diabatic coupling, and $\omega_{j,\alpha}$ and $c_{j,\alpha}$ are the frequency and the coupling strength of the $j$th mode of bath $\alpha$. Both baths have a Debye spectral density
    \begin{align}
        J_\alpha(\omega) = \frac{\lambda_\alpha\omega\Omega_{\alpha}}{2(\omega^2+\Omega_{\alpha}^2)},
    \end{align}
    where $\lambda_\alpha$ and $\Omega_{\alpha}$ are the Marcus reorganization energy and characteristic frequency of bath $\alpha$.
    The bath correlation functions can be evaluated analytically to give\cite{Carmichael1}
    \begin{subequations}\label{eq:Gamma}\begin{align}
        \Re[\Gamma_\alpha(\omega)] &= \frac{J_\alpha(\omega)}{1-\eu{-\beta\omega}}
        \\
        \Im[\Gamma_\alpha(\omega)] &= \frac{1}{\pi}\fint_{-\infty}^{\infty} \frac{J_\alpha(\omega')}{1-\eu{-\beta\omega'}}\frac{\dd\omega'}{\omega-\omega'}\,, \label{eq:ImGamma}
    \end{align}\end{subequations}
    where
    Eq.~\eqref{eq:ImGamma} is defined by its Cauchy principal value. While, in this case, the system--bath coupling is not explicitly position-dependent, the jumping probabilities and the Lamb shift [Eq.~\eqref{eq:gamma_xi}] do depend on position as they are defined in terms of the instantaneous adiabatic states. 
    
    The system is initialized in diabatic state $\ket{a}$, and the classical modes are sampled from the Boltzmann distribution of the uncoupled classical bath
    \begin{equation}
        \rho_0(\bm{q},\bm{p}) = \prod_{j}\frac{\beta\omega_{j,\mathrm{c}}}{2\pi}\exp\left[-\beta\left(\frac{p_{j,\mathrm{c}}^2}{2} + \frac{1}{2}\omega_{j,\mathrm{c}}^2 q_{j,\mathrm{c}}^2\right)\right].
    \end{equation}

\begin{figure}
        \includegraphics[width=\linewidth]{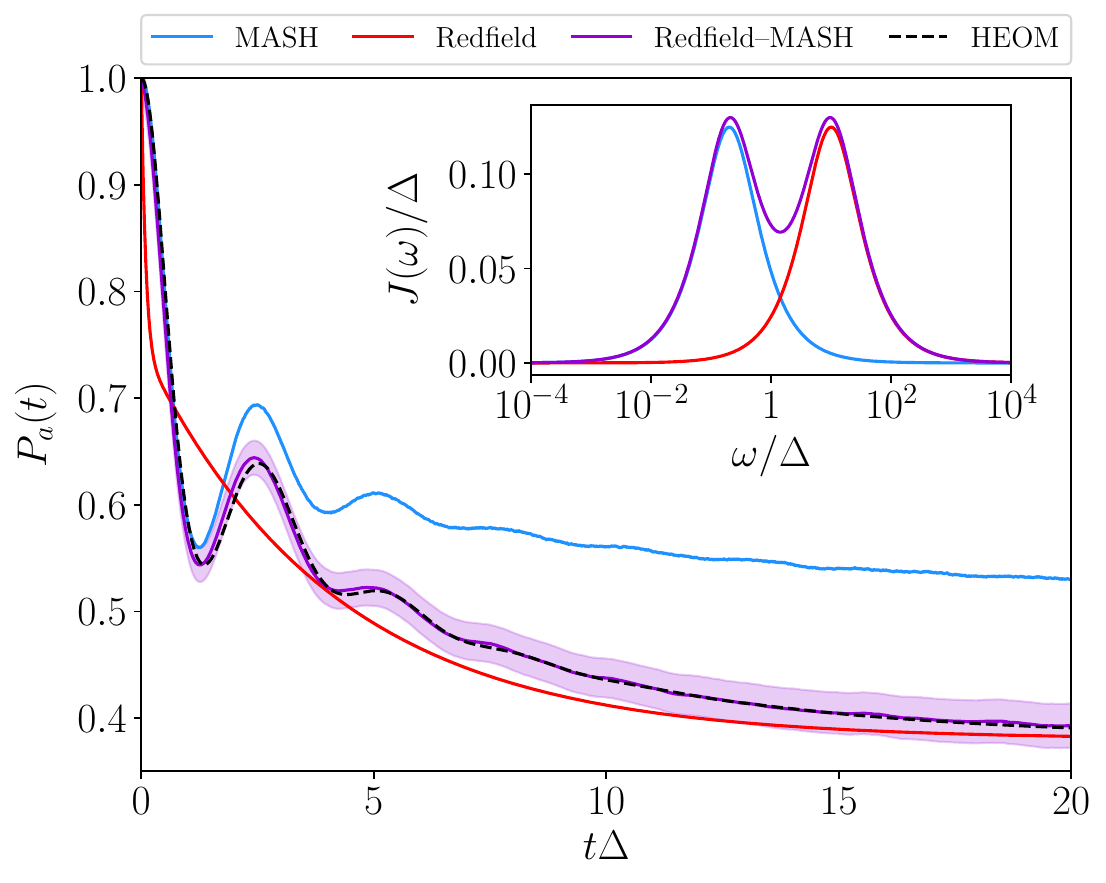}
        \caption{The population of the diabatic $\ket{a}$ state as a function of time using MASH (blue), secular Redfield theory (red), hybrid Redfield--MASH (purple), and HEOM (dashed). 20$\sigma$ error bars indicate a 95\% confidence interval for $10^4$ trajectories of the hybrid method. Inset: The spectral density of the slow bath (blue), the fast bath (red) and the total spectral density (purple) of the spin--boson system.}
        \label{fig:sb_results}
    \end{figure}

    We chose the parameters of the model such that the classical bath can be simulated well with MASH and the quantum bath using secular Redfield theory: $\varepsilon / \Delta = 1$, $\beta\Delta = 0.25$, $\Omega_{\mathrm{c}} / \Delta = 0.2$, $\Omega_{\mathrm{q}} / \Delta = 10$, $\lambda_\mathrm{c} / \Delta  = \lambda_\mathrm{q} / \Delta = 0.5$. The spectral density of the baths is illustrated in the inset of Fig.~\ref{fig:sb_results}. The classical bath has the same parameters as a single-bath model used in Ref.~\onlinecite{MASH}, where it was shown that MASH gives accurate results. The quantum bath is described well by Redfield theory, as it is weakly coupled and Markovian over the timescale relevant to the system. On the other hand, the classical bath is too slow to be described well by Markovian master equations, and the quantum bath is cold ($k_\mathrm{B} T=\beta^{-1} \ll \Omega_{\mathrm{q}}$) so that quantum effects such as zero-point energy cannot be ignored. 
    
    As described at the end of Sec.~\ref{sec:redfield}, secular Redfield theory (which underpins the hybrid method) is only formally valid for time-dependent Hamiltonians which change slowly compared to the bath relaxation time and whose intrinsic system dynamics evolve on a short timescale. The first requirement is satisfied in this case as the time-dependence of the Hamiltonian in our hybrid method follows the slow bath. The second requirement 
    holds because the slowest intrinsic timescale of the system 
    (which is inversely proportional to the adiabatic energy gap near the avoided crossing) remains shorter than the overall relaxation timescale.

    We simulated the diabatic populations as a function of time in the two-bath problem using the pure MASH method, 
    secular Redfield theory, and our hybrid method. To provide an exact quantum-mechanical benchmark, we employed HEOM,\cite{Tanimura2020HEOM} as implemented in the \texttt{pyrho} package.\cite{pyrho} In MASH and our hybrid method, the diabatic populations are calculated using Eq.~\eqref{eq:diabatic_population}.  We sample the initial spin vector uniformly from the surface of the Bloch sphere.  
    
    For the hybrid method, the classical bath is discretized into 200 modes identically to the original MASH paper,\cite{MASH} and for the pure MASH simulation of the two-bath model, we discretize both baths into 200 modes and sample 
    both from Boltzmann distributions, whereas pure Redfield theory [Eq.~\eqref{eq:redfield_secular_me}] treats both baths implicity via the Fourier transforms of the bath correlation functions, Eq.~\eqref{eq:Gamma}. 
    The results were converged with respect to the timestep, number of trajectories and number of bath modes. In particular, we used timesteps of $0.0005\Delta$ for the MASH simulations and $0.01\Delta$ for the hybrid simulations.  As expected, the classical treatment of the high-frequency bath requires significantly smaller timesteps and therefore increased computational cost, while the hybrid treatment of the problem can employ longer timesteps and thus has a similar computational cost to a MASH simulation of the low-frequency bath only.
    Note that better integrators could probably be designed for this problem using exact updates of the harmonic degrees of freedom, or by replacing the classical bath with a generalized Langevin equation.\cite{Zwanzig}
    For full convergence, we employed $10^6$ trajectories for both MASH and the hybrid methods. However, we present error bars of 20$\sigma$ for the hybrid method, which is equivalent to the 95\% confidence interval that would be obtained when using $10^4$ trajectories, which indicates that in practice, reasonable results can be obtained with far fewer trajectories than we have used.
    
    Figure~\ref{fig:sb_results} shows the time-evolved population of the diabatic $\ket{a}$ state. Redfield theory can accurately model the fast bath and MASH can accurately model the slow bath, but they fail in describing the system coupled to both baths simultaneously,
    as the high-frequency bath is too quantum for MASH and the low-frequency bath is too non-Markovian for Redfield theory.
    On the other hand, our hybrid approach is in excellent agreement with the fully quantum--mechanical HEOM benchmark and tends to the same equilibrium distribution as Redfield theory and HEOM (within the limit that the low-frequency bath can be well approximated as classical).
    
    One might assume that using Wigner distributions with pure MASH could improve its description of the quantum bath.  However, the zero-point energies of the high-frequency modes sampled with a Wigner distribution would leak into the system as well as the lower-frequency modes of the bath.\cite{Habershon2009water} Depending on how many modes we use to discretize the bath, this can heat up the system to arbitrarily high temperatures, leading to the incorrect result $P_a(t\rightarrow\infty)\rightarrow\half$.\cite{thermalization}
    One could try to control the unphysical flow of zero-point energy,\cite{Varandas1994ZPE,miller1989simple,bowman1989method,BenNun1996ZPE} although such tricks are typically not rigorous and can in some cases lead to even worse results results.\cite{Guo1996ZPE}
    So-called ``focused sampling'' in action--angle variables to give each bath mode exactly its ZPE\cite{ZPE1} will not help here either as in an ergodic system, the long-time limit depends only on the total energy and not on the fine details of the initial conditions.
    The hybrid Redfield--MASH approach, however, offers a practical scheme to fix the problems of zero-point energy leakage in nonadiabatic trajectory simulations, as long as the high-frequency modes have the required properties (weakly coupled and Markovian).

    \subsection{Spontaneous emission}\label{sec:spontaneous_emission}
    Master equations such as Eq.~\eqref{eq:redfield_secular_me} are widely used in quantum optics to model the interaction of light with matter, as the underlying Markovian approximations are well satisfied in many relevant cases. Following Ref.~\onlinecite{OpenQuantum}, the zero-temperature photon bath (which describes the vacuum fluctuations of the electromagnetic field) can be modeled in the dipole approximation using
    \begin{subequations}
    \begin{align}
        \Hsb  &= -\hat{\bm{\mu}}\cdot\hat{\bm{E}}, \\
        \Re[\Gamma^{(\rm vac)}(\omega)] &= \begin{cases}
            \frac{2\omega^3}{3c^3} & \omega > 0\\
            0 & \omega \leq 0 
        \end{cases}, \\
        \gamma^{(\rm vac)}_- &= 2|\braket{\Phi_0|\hat{\bm{\mu}}|\Phi_1}|^2\Re[\Gamma^{(\rm vac)}(\omega_\mathrm{S})] ,\label{eq:gamma_thermal}
    \end{align}
    \end{subequations}
    where $\hat{\bm{\mu}}$ is the dipole operator of the system, 
    $\hat{\bm{E}}$ is the quantum electric-field operator, $c$ is the speed of light, $\Re[\Gamma^{(\rm vac)}(\omega)]$ is the Fourier transform of the zero-temperature photon bath correlation function, $\gamma^{(\rm vac)}_-$ is the spontaneous-emission decay rate to the vacuum photon field.
    No other jumps are possible in this case, as the photon bath at zero-temperature is not occupied and thus cannot excite the system via $\hat{\sigma}_+$ jumps. Additionally, there are no zero-frequency photons to initiate $\hat{\sigma}_z$ jumps.
    Note that the Lamb shift is not well defined in this case without introducing relativistic corrections.\cite{lamb-shift-div} In practice, this shift corresponds to a renormalization of the system energy and is commonly absorbed in the system Hamiltonian or neglected altogether.\cite{Rivas2012} Therefore, we follow standard practice and ignore the Lamb shift in our treatment of this problem.

    Spontaneous emission is typically slow compared to the ultrafast timescale of nonadiabatic relaxation. Its rate can however be enhanced when the molecules are placed in an optical cavity, referred to as the Purcell effect. For a two-state system, the master equation at zero temperature can be written as\cite{Carmichael2}
    \begin{align}
        \dot{\hat{\rho}} &= -i\frac{\omega_\mathrm{S}}{2}\left[\hat{\sigma}_z, \hat{\rho}\right] - i\omega_\mathrm{cav}\left[\hat{a}^\dagger\hat{a}, \hat{\rho}\right] + g\left[\hat{a}^\dagger\hat{\sigma}_- - \hat{a}\hat{\sigma}_+, \hat{\rho}\right] \nonumber \\
        &\quad + \gamma^{(\rm vac)}_-\left(\hat{\sigma}_-\hat{\rho}\hat{\sigma}_+ - \tfrac{1}{2}\{\hat{\sigma}_+\hat{\sigma}_-,\hat{\rho}\} \right) \nonumber \\
        &\quad + 2\kappa \left(\hat{a}\hat{\rho}\hat{a}^\dagger - \tfrac{1}{2}\{\hat{a}^\dagger\hat{a},\hat{\rho}\}\right),
    \end{align}
    where $\omega_\mathrm{cav}$ is the cavity frequency, $\hat{a}$ and $\hat{a}^\dagger$ are the cavity-mode annihilation and creation operators, $g$ is the system--cavity coupling, and $\kappa$ is the cavity-mode damping rate.
    
    In this section, we will demonstrate how our model can simulate cavity-enhanced emission in the \textit{leaky-cavity limit} ($\kappa \gg \half\gamma^{(\rm vac)}_-, \kappa \gg g$), where we assume any photon emitted to the cavity is dissipated before it can be reabsorbed by the system. This allows us to adiabatically eliminate the cavity mode from the system Hamiltonian and include it as part of the photon bath. With this assumption, we can reduce the problem to a molecule coupled to a single photonic bath, described using Eq.~\eqref{eq:redfield_secular_me}, for which the cavity-enhanced decay rate is\cite{Carmichael2}
    \begin{align} \label{eq:cavity}
        \gamma_- &= \gamma^{(\rm vac)}_-\left(1 + 2C_1\frac{\kappa^2}{\kappa^2 + (\omega_\mathrm{cav}-\omega_\mathrm{S})^2}\right),
    \end{align}
    where the spontaneous-emission enhancement factor is
    \begin{equation}
        2C_1\equiv \frac{g^2}{\kappa\Re[\Gamma^{(\rm vac)}(\omega_\mathrm{S})]}.
    \end{equation}
    In this description, the cavity mode and the background photon field are treated simultaneously in the bath correlation functions and decay rates.

\begin{figure}
        \includegraphics[width=\linewidth]{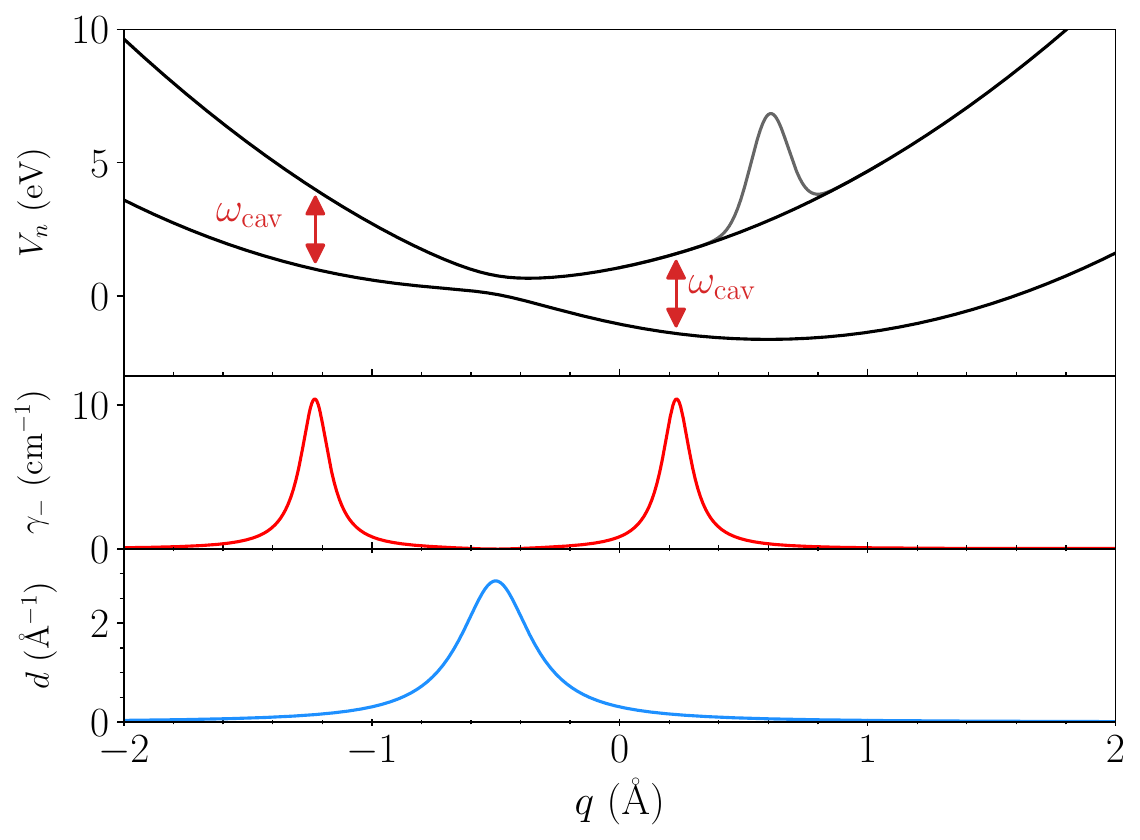}
        \caption{The adiabatic potentials of Eq.~\eqref{eq:electron_transfer} and the corresponding cavity-enhanced decay rates and nonadiabatic couplings. Red arrows show the energy gap at resonance with the cavity. The initial wavepacket is also plotted in gray.}
        \label{fig:emission_pot}
    \end{figure}

    To demonstrate that our hybrid method can accurately model the competition between radiative and non-radiative decay of a molecule in an excited state, we present the following one-dimensional model of a two-state molecule inside a cavity in the diabatic basis
    \begin{align}
        \Hs &= \frac{p^2}{2m} + \tfrac{1}{2}m\omega_0^2q^2 + \begin{pmatrix}
            \varepsilon + \zeta q & \Delta \\
            \Delta & -\varepsilon - \zeta q
        \end{pmatrix}, \label{eq:electron_transfer}
    \end{align}
    where we set $\omega_0 = 300\,\text{cm}^{-1}$, $\varepsilon=1\,\text{eV}$, $\Delta = 0.35\,\text{eV}$, $\zeta = 2\,\text{eV}/\text{\AA}$, and $m = 10\,\text{amu}$. The adiabatic potentials (obtained by diagonalization) are displayed in Fig.~\ref{fig:emission_pot}. 
    Note that even though the diabats are harmonic, the adiabatic potentials are not, making this a non-trivial problem for quantum--classical dynamics.
    The dipole operator in the diabatic basis is 
    \begin{align}
        \hat{\mu} &= \begin{pmatrix}
            0 & \mu_{ab} \\
            \mu_{ab} & 0
        \end{pmatrix},
    \end{align}
    where $\mu_{ab} = 5\,\text{D}$. \footnote{We use a purely off-diagonal dipole moment operator for simplicity, as diagonal components in the diabatic representation will have only a minor effect on the transition dipole moment measured in the adiabatic representation,
    at least in the relevant regions on resonance with the cavity.  Nonetheless, the method can easily be applied to the more general case, where the trace of the dipole operator should be absorbed in the potential energy surfaces, and the traceless remainder can be treated using our hybrid method.}
    The cavity parameters are $g = 110\,\text{cm}^{-1}$, $\kappa = 2200\,\text{cm}^{-1}$, and $\omega_\mathrm{cav} = 3\,\text{eV}$. Figure~\ref{fig:emission_pot} shows the nonadiabatic coupling and 
    the cavity-enhanced decay rate [Eq.~\eqref{eq:cavity}], which depends on $q$ through $\omega_\mathrm{S}$ and the transition dipole moment between the adiabatic states. 
    On resonance ($q \approx 0.23$ or $-1.23$\,\AA), the enhancement is $2C_1(\omega_\mathrm{cav}) \approx 18637$, which as we shall show, ensures that the two modes of decay from the excited state (nonadiabatic coupling and spontaneous emission) have similar rates.

    The nuclei are initialized in the vibrational ground state of 
    diabat $\ket{b}$ and excited to the upper adiabatic state (as illustrated in Fig.~\ref{fig:emission_pot}). In the trajectory simulations, the initial nuclear Wigner distribution is thus given by
    \begin{equation}
        \rho_0(q,p) = \frac{1}{\pi}\exp\left[-\frac{p^2}{m\omega_0} - m\omega_0 \left(q - \frac{\zeta}{m\omega_0^2}\right)^2\right].
    \end{equation}
    We measure the time-dependent adiabatic population, $P_1$(t), following the improved procedure of Eq.~\eqref{eq:sz_plus_1}. 
    
\begin{figure}
        \includegraphics[width=\linewidth]{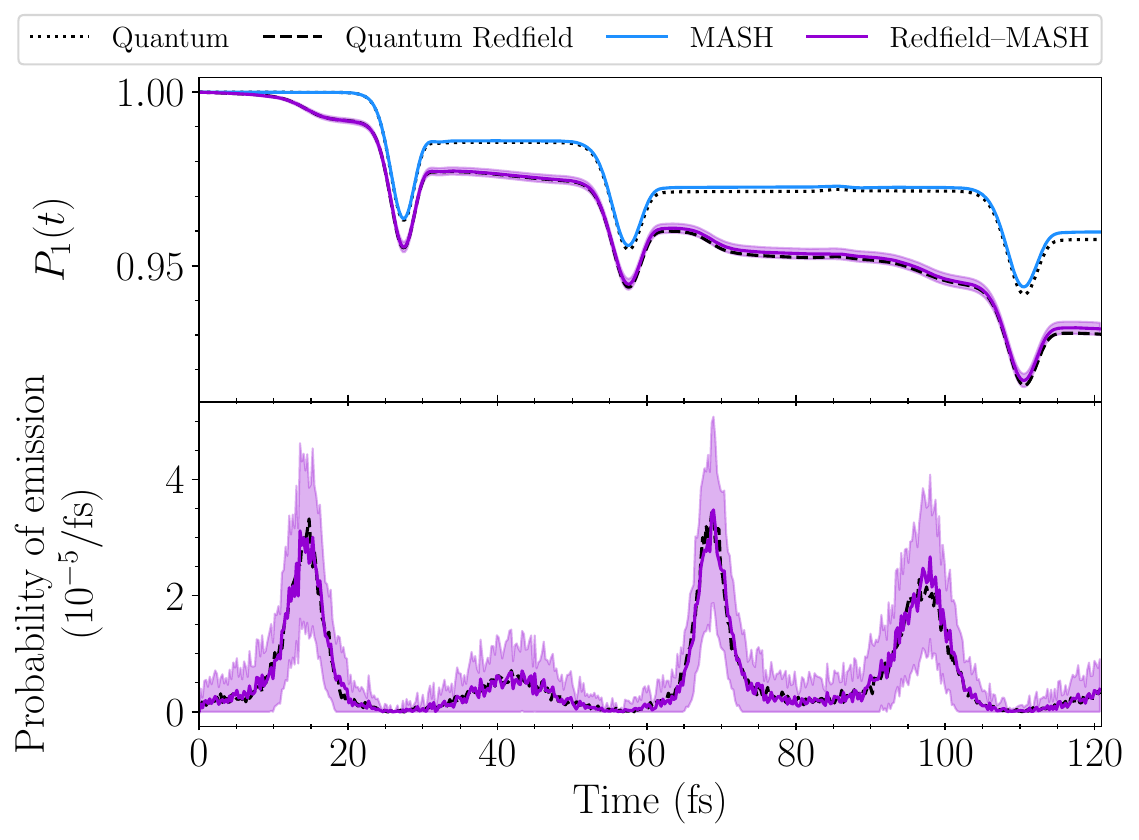} 
        \caption{Top: Time-dependent population of the upper adiabatic state. The quantum and MASH calculations are for an isolated system, and the quantum Redfield and hybrid Redfield--MASH results are for a coupled system--cavity simulation. Bottom: Probability of photon emission per unit time for both Redfield methods. $2\sqrt{10}\sigma$ error bars indicate a 95\% confidence interval for $10^5$ trajectories of the hybrid method.}
        \label{fig:Pt_leaky}
    \end{figure}

     In Fig.~\ref{fig:Pt_leaky}, we compare the results from our hybrid method to the quantum Redfield population dynamics obtained using the stochastic formulation of Sec.~\ref{sec:stochastic} propagated on a grid using the split-operator method.
     For comparison, we also show results from pure MASH and pure quantum simulations of the isolated system.
     We used $10^6$ trajectories in each case to fully converge the weak radiative decay and the photon emission probabilities. In practice it is not necessary to run so many trajectories to obtain reasonable results, and Fig.~\ref{fig:Pt_leaky} shows error bars of $2\sqrt{10}\sigma$ 
     for the hybrid method, which is equivalent to the 95\% confidence interval that would result from using only $10^5$ trajectories. All methods use timesteps of 10\,a.u. Note that without the second-order algorithms developed in Ref.~\onlinecite{MASHEOM}, we would require timesteps two orders of magnitude shorter in order to accurately capture the small nonadiabatic population transfer in this model.
    
    Figure~\ref{fig:Pt_leaky} plots the population of the excited state obtained from both pure MASH (without any coupling to the cavity or the vacuum) and our hybrid method.
    Here, one can clearly see the two distinct decay channels: the nonadiabatic transition due to the coupling between the nuclear and electronic degrees of freedom, and the radiative decay via spontaneous emission into the photon bath. Pure MASH captures only the nonadiabatic transitions of the isolated system, as seen in the drops in the adiabatic population at roughly 27, 57 and 111\,fs, where the wavepacket passes through the avoided crossing near $q\approx-0.5$\,\AA.
    The hybrid method additionally captures spontaneous emission to the cavity mode around 14, 43, 69 and 98 fs, when the wavepacket passes through the regions resonant to the cavity mode (indicated by the red arrows in Fig.~\ref{fig:emission_pot}). 

    One could in principle attempt to simulate the full system and (discretized) photon bath using MASH\@. However, this approach would face the same problems as the MASH simulations of the quantum bath in Sec.~\ref{sec:spin-boson}. The relevant modes of the photon bath have very high frequencies, and as we want to simulate the bath at zero temperature, classical methods will generally fail. In particular, initializing with a Wigner distribution will lead to zero-point energy leakage, which will unphysically heat up the system.

    Secular Redfield theory works well for this problem as both the Markovian and the secular approximation are valid. In particular, the total relaxation time is longer than the bath relaxation time $\sim\kappa^{-1}$ and the fastest intrinsic system timescale $\sim\Delta^{-1}$. Note that the quantum Redfield approach used as a reference in this section employs a nuclear wavefunction discretized on a grid.
    This approach is similar to the Redfield description of the dissipative dynamics of molecules developed in Ref.~\onlinecite{quantum-redfield}, except that in our simulation, we neglect the extremely weak coupling of the photon bath to the vibrational states of the molecule. 
    Both approaches scale exponentially with the number of nuclear degrees of freedom and are computationally intractable for molecular systems with more than a few atoms. In contrast, our hybrid approach, which gives similar results, treats the nuclei using classical trajectories and is therefore scalable to systems with many nuclear degrees of freedom. This makes our method suitable for realistic molecular simulations to provide mechanistic insight into the dynamics of fluorescent molecules.

\section{Conclusions}
    We have developed a hybrid quantum--classical method that unifies trajectory-based nonadiabatic molecular dynamics with a dissipative master-equation description of quantum baths. By combining the deterministic dynamics of MASH with the stochastic dynamics of the unravelled Redfield theory, this approach allows for a simultaneous treatment of a two-level system interacting with classical degrees of freedom and a quantum environment within a unified framework.

    Application to the spin--boson model with two baths demonstrates that the hybrid method is in good agreement with exact HEOM dynamics. 
    This is in contrast to pure MASH, which cannot correctly account for zero-point energy in the high-frequency modes, or pure Redfield theory, which cannot correctly describe the non-Markovian slow bath modes. In another setting, the hybrid method naturally describes fluorescence dynamics and cavity-enhanced spontaneous emission, demonstrating its applicability to quantum-optics problems. In fact, this approach offers a powerful framework which is not limited to these two problems. For instance, it would also be applicable to non-thermal quantum environments such as squeezed states. Additionally, it could describe the fermionic baths that are encountered when modeling molecules interacting with metal surfaces, while avoiding the difficulties with developing a fermionic mapping approach.\cite{Miller1986fermions,Li2012fermions,Montoya2018fermions,sun2021bosonic}

    The trajectory-based hybrid approach we present in this paper retains the scalability and the interpretability of surface-hopping methods, while allowing for dissipative dynamics through a master-equation treatment of the environment, as long as one can successfully identify which modes to treat classically and which to include in the quantum bath. The validity of the method is currently limited to weakly-coupled Markovian quantum baths, as dictated by the approximations of the secular Redfield method.
    Further work will focus on relaxing these approximations, using non-secular Redfield theory, allowing for non-Markovian environments, and going beyond a perturbative treatment of the system--bath coupling. Together with on-the-fly ab initio electronic-structure calculations, this will enable realistic simulations of molecular systems in quantum environments. 

\section*{Conflict of Interest Statement}
The authors have no conflicts to disclose.

\section*{Data Availability}
The data that support the findings of this study are available within the article.

\appendix

\section{Alternative MASH weighting factors}\label{ap:weighting}

\subsection{Non-unitary quantum dynamics}{\label{ap:weighting_cp}}

The original MASH weighting factors introduced in previous work were chosen because it could be proved that they capture the correct unitary quantum dynamics along a predetermined classical pathway, $\bm{q}_t$.\cite{MASH,MASHreview} 
In this section, we justify the alternative set of weighting factors that do not require the dynamics to be unitary and are therefore also applicable to non-unitary quantum dissipative dynamics.

In particular, the coherence--population weighting factor $W_{\rm CP}$, differs in our method [Eq.~\eqref{eq:mash_weights}] compared to the original MASH formulation. In justifying $W_{\rm CP}=2$ in the original MASH method\cite{MASH} (see in particular the Supplementary Information of Ref.~\onlinecite{MASHreview}), we employed a variable transform so as to perform the integral over the spins at time $t$, in which we assumed $\bm{S}(t)=\mathsf{U}(t)\bm{S}$, where $\mathsf{U}(t)$ is an orthogonal rotation matrix (as rotations of the spin vector correspond to unitary dynamics in the Hilbert space).\cite{MahlerBook} We cannot make this assumption in our problem as quantum dissipative dynamics is not unitary.

Here, we do not yet consider the full dissipative equations of motion, but assume a linear dynamics 
$\dot{\hat{\sigma}}_\mu(t) = \sum_\nu \Omega_{\mu\nu}(t) \hat{\sigma}_\nu(t)$, which may or may not be unitary.
To find a more general set of weighting factors, we first consider a coherence--population correlation function such as $C_{xz}(t)=\Tr\left[\hat{\sigma}_x\hat{\sigma}_z(t)\right]$
with the time derivative $\dot{C}_{xz}(t)=\sum_\nu \Omega_{z\nu} C_{x\nu}(t)$.

The MASH equivalent of this correlation function is defined as
\begin{align}
    C^{\rm MASH}_{xz}(t) &= \int S_x \, W_{\rm CP}(t) \sgn S_z(t)\,\dd \bm{S},
\end{align}
and we want to choose $W_{\rm CP}(t)$ such that 
the MASH correlation function matches the quantum result, $C_{xz}(t)$.
It is easy to show that the mapping guarantees that all correlation functions match at the initial time, $C_{\mu\nu}(0)=C_{\mu\nu}^\mathrm{MASH}(0)$.
Therefore, it is sufficient to show that their first derivatives match to ensure that they are equal for all time. If we define the spin equations of motion in MASH as 
$\dot{S}_\mu(t) = \sum_\nu \Omega_{\mu\nu}(t) S_\nu(t)$,
the correct behavior is achieved using $W_{\rm CP}(t)=3|S_z(t)|$:
\begin{align}
    C^{\rm MASH}_{xz}(t) = \int S_x 3 |S_z(t)|\sgn{S_z(t)}\,\dd \bm{S} = \int 3 S_x S_z(t)\,\dd \bm{S} ,
\end{align}
which (using $W_{\rm CC}=3$) leads to
\begin{align}
    \dot{C}^{\rm MASH}_{xz}(t)
    &= \int 3S_x \, \dot{S}_z(t)\,\dd \bm{S} \nonumber\\
    &= \int 3S_x \, [\Omega_{zx}S_x(t) + \Omega_{zy}S_y(t) + \Omega_{zz}S_z(t)]\,\dd \bm{S} \nonumber\\
    &= \sum_{\nu}\Omega_{\nu}C^{\rm MASH}_{x\nu}(t),
\end{align}
as required.

The other weighting factors in Eq.~\eqref{eq:mash_weights}, which are the same as the original MASH method, can be justified in a similar manner through population--population, population--coherence and coherence--coherence correlation functions (e.g.\ $C_{zz}, C_{zx}, C_{xx}$).

This confirms that the new weighting factors [Eq.~\eqref{eq:mash_weights}] obey the same desirable property as the original MASH, in that they recover the correct quantum dynamics along a predetermined classical pathway. However, the new version is more general as it allows for non-unitary quantum dynamics.
Appendix~\ref{ap:proof} demonstrates how these weighting factors are also appropriate for the nonlinear equations of motion presented in Eq.~\eqref{eq:spin_evol} for the hybrid method.

\subsection{Connection to the QCLE} \label{ap:QCLE}

Another important proof which underpins the validity of the MASH approach is that its dynamics are in agreement with the QCLE\cite{Kapral2015QCL} to first-order in time.\cite{MASH,MASHreview}
Reproducing the correct electronic dynamics (where we account for the time-dependent weighting functions) 
is necessary but not sufficient to show this, as here we also include the back reaction of the spin dynamics on the classical degrees of freedom.
In order to determine if the new weighting factors also share this property, 
we need to consider whether the following equalities hold [Eq.~(A15) of Ref.~\onlinecite{MASHreview}]:\footnote{Note the typo in Eq.~(A15b) of the Supplementary Material of Ref.~\onlinecite{MASHreview}, in which $S_x$ is missing.}
\begin{subequations}\begin{align}
    \int \mathcal{A}_\mu W_{\mu\nu}\pder{\mathcal{B}_\nu}{p_j}\sgn(S_z)\,\dd\bm{S} &= \Tr\left[\hat{\mathcal{A}}_\mu\frac{1}{2}\left[\pder{\hat{\mathcal{B}}_\nu}{p_j},\hat{\sigma}_z\right]_+\right],\label{eq:qcle_requirement1}
    \\
    2\int \mathcal{A}_\mu W_{\mu\nu}\pder{\mathcal{B}_\nu}{p_j} S_x \delta(S_z)\,\dd\bm{S} &= \Tr\left[\hat{\mathcal{A}}_\mu\frac{1}{2}\left[\pder{\hat{\mathcal{B}}_\nu}{p_j},\hat{\sigma}_x\right]_+\right],\label{eq:qcle_requirement2}
\end{align}\label{eq:qcle_requirement}\end{subequations}
where $\mu,\nu\in\{I,x,y,z\}$. 
It is easy to show that in most cases, these equations are obeyed.  However, Eq.~\eqref{eq:qcle_requirement2} does not hold with the new weighting factor $W_\mathrm{CP}=3|S_z(t)|$ 
in the special case of $\mu\in\{x,y\}$ and $\nu=I$. However, this term is only relevant when calculating correlation functions starting in an electronic coherence and measuring an operator of the nuclear momentum at a later time.
Except for this rare case, which is not relevant for any case in this work, 
MASH with the new weighting factors still obeys the QCLE to first-order in time. 
This ensures that the correlation functions are not only correct at $t=0$ but that they have the correct initial slope.

\section{Derivation of the electronic equations of motion}\label{ap:proof}

    In this appendix, we derive a PDP equivalent to Eq.~\eqref{eq:redfield_secular_me_time} for an ensemble of spin vectors within the MASH framework (in contrast to the ensemble of wavefunctions used in Sec.~\ref{sec:stochastic}). Any operator $\hat{O}$ can be mapped onto a function of spin vectors using $O(\bm{S}) = {\rm tr}[\hat{O} \, \hat{w}(\bm{S})]$, where the MASH kernel is
    \begin{align}
        \hat{w}(\bm{S}) &= \tfrac{1}{2}(\hat{I} + \hat{\sigma}_xS_x + \hat{\sigma}_yS_y + \hat{\sigma}_z\sgn S_z).\label{eq:mash_kernel}
    \end{align}
    In this way, a density matrix $\hat{\rho} = \tfrac{1}{2}(\hat{I} + \hat{\sigma}_x\rho_x + \hat{\sigma}_y\rho_y + \hat{\sigma}_z\rho_z)$ is mapped to $\rho(\bm{S}) = {\rm tr}[\hat{\rho}\,\hat{w}(\bm{S})] = \tfrac{1}{2}(1 + \rho_xS_x + \rho_yS_y + \rho_z\sgn{S_z})$.
	
    We need to define the deterministic and stochastic parts of the spin evolution to reproduce the desired density-matrix evolution. The time-evolved spin vector $\bm{S}(t + \dd t) = \bm{S}'$ can be considered to come from the distribution over non-normalized spin vectors
    \begin{subequations}\begin{align}
        \bm{S}' &\sim (1 -\Sigma(t)\dd t)\,\delta(\bm{S}' - \bm{S}(t) - \dot{\bm{S}}\dd t) \nonumber\\
        &\quad+ \mathcal{P}_+(t)f_+(\bm{S}')\,\dd t +  \mathcal{P}_-(t)f_-(\bm{S}')\,\dd t \nonumber \\
        &\quad+  \mathcal{P}_z(t)f_z(\bm{S}')\,\dd t, \label{eq:evol_dist}\\
        \Sigma(t) &= \mathcal{P}_+(t) + \mathcal{P}_-(t) + \mathcal{P}_z(t) ,
    \end{align}\end{subequations}
    where $\mathcal{P}_\pm(t)\dd t$ and $\mathcal{P}_z(t)\dd t$ are the jump probabilities, and $f_\pm(\bm{S}')$ and $f_z(\bm{S}')$ are distributions from which the spin is sampled after each jump.
    The first term in Eq.~\eqref{eq:evol_dist} corresponds to deterministic evolution with the probability $1-\Sigma(t)\dd t$ that no jump occurs.

    To derive our mapped version of the stochastic master equation, we propose the following ansatzes for the post-jump distributions
    \begin{subequations}\begin{align}
        f_\pm(\bm{S}') &= h(\pm S'_z) \, \delta(|\bm{S}'|^2-1) , \\
        f_z(\bm{S}') &= \delta(S'_x+S_x)\,\delta(S'_y+S_y)\,\delta(S'_z-S_z),
    \end{align}\end{subequations}
    and for the deterministic evolution, 
    \begin{subequations}
        \begin{align}
            \dot{S}_x &= -\omega_{\rm LS}(t) S_y + 2\tau(t)S_z + \mathcal{G}_x(\bm{S}), \\
            \dot{S}_y &= \omega_{\rm LS}(t) S_x + \mathcal{G}_y(\bm{S}), \\
            \dot{S}_z &= -2\tau(t)S_x + \mathcal{G}_z(\bm{S}).
        \end{align}
    \end{subequations}
    Our goal is then to derive the probabilities, $\mathcal{P}_\pm(t), \mathcal{P}_z(t)$, and dissipative terms, $\mathcal{G}_x(\bm{S}), \mathcal{G}_y(\bm{S}), \mathcal{G}_z(\bm{S})$, such that the time evolution of the averaged ensemble of spin vectors reproduces the quantum dynamics of the reduced density matrix [Eq.~\eqref{eq:redfield_secular_me_time}]. 

\subsection{Weighting factors}\label{ap:proof_weight}

    When a stochastic jump operator is applied, the spin vector is resampled. In this section, we show how the weighting factors need to be modified after each jump. Inspired by the quantum-jump procedure in the original MASH method,\cite{MASH} we consider the evolution of a density matrix $\hat{\rho}(t_0) = \hat{O}$ to $\hat{\rho}(t_1)$. In MASH, this is given by
    \begin{align}
        \hat{\rho}(t_1) = \frac{1}{2}\bigg[&\int\dd \bm{S} \, O(\bm{S})W_{O\mathrm{P}}(t_1)\big(\hat{I} + \hat{\sigma}_z \sgn S_z(t_1)\big) \nonumber \\
        +& \int\dd \bm{S} \, O(\bm{S})W_{O\mathrm{C}}(t_1)\big(\hat{\sigma}_x S_x(t_1) + \hat{\sigma}_y S_y(t_1)\big) \bigg]. \label{eq:rho_mash_t}
    \end{align}
	
    At time $t_1$, one of the jump operators is applied to the system, for example $\hat{\sigma}_- = \ketbra{\Phi_0}{\Phi_1}$. This effectively measures a population in the upper state and resets the density matrix in the lower state, thus transforming it from $\hat{\rho}(t_1)$ to 
    \begin{align}
        \hat{\rho}' &=\hat{\sigma}_-\hat{\rho}(t_1)\hat{\sigma}_-^\dagger = \int \dd \bm{S} \, O(\bm{S}) W_{O\mathrm{P}}(t_1) h(S_z(t_1))\ketbra{\Phi_0}{\Phi_0},
    \end{align}
    which we will then map to a spin-vector on the Bloch sphere as 
    \begin{align}
        \rho'(\bm{S}') &= {\rm tr}\left[\hat{\rho}' \hat{w}(\bm{S}')\right] = \int \dd \bm{S} \, O(\bm{S})W_{O\mathrm{P}}(t_1) h(S_z(t_1))h(-S'_z).
    \end{align}
    We then evolve this mapped density matrix to time $t>t_1$,
    \begin{align}
        \hat{\rho}(t) &= \frac{1}{2}\iint \dd \bm{S} \, \dd \bm{S}' \, O(\bm{S})W_{O\mathrm{P}}(t_1) h(S_z(t_1))h(-S'_z)\nonumber \\
        &\quad\times \left[ W_{\mathrm{P}\mathrm{P}}(t)\big(\hat{I} + \hat{\sigma}_z \sgn S'_z(t)\big) \right. \nonumber \\
        &\quad\left.+ W_{\mathrm{PC}}(t)\big(\hat{\sigma}_x S'_x(t) + \hat{\sigma}_y S'_y(t)\big) \right] \nonumber\\
        &= \frac{1}{2}\iint \dd \bm{S} \, \dd \bm{S}' \, O(\bm{S})\left[ W_{O\mathrm{P}}^{(1)}(t;t_1)\big(\hat{I} + \hat{\sigma}_z \sgn S'_z(t)\big) \right. \nonumber \\
        &\left.\quad+ W_{O\mathrm{C}}^{(1)}(t;t_1)\big(\hat{\sigma}_x S'_x(t) + \hat{\sigma}_y S'_y(t)\big)\right],
    \end{align}
    where we define the new weighting factors after one hop as
    \begin{subequations}\begin{align}
        W_{O\mathrm{P}}^{(1)}(t;t_1) &= W_{O\mathrm{P}}(t_1) h(S_z(t_1)) h(-S'_z)W_{\mathrm{P}\mathrm{P}}(\bm{S}'(t)), \\
        W_{O\mathrm{C}}^{(1)}(t;t_1) &= W_{O\mathrm{P}}(t_1) h(S_z(t_1))h(-S'_z)W_{\mathrm{PC}}(\bm{S}'(t)).
    \end{align}\end{subequations}

    Note that the weighting factors for $\hat{\sigma}_\pm$ jumps measure $h(\mp S_z(t_1))$ using the spin vector immediately before the jump. Therefore, in order to avoid calculating trajectories with vanishing weighting factors, we include the Heaviside term in the jumping probabilities $\mathcal{P}_\pm(t)\propto h(\mp S_z(t))$.
    The factor of $h(\pm S_z')\,\delta(|\bm{S}'|^2 - 1)$ is accounted for by only sampling the new spin vector from the appropriate hemisphere.
    Generalizing this procedure for multiple jumps of either $\hat\sigma_+$ or $\hat\sigma_-$, we obtain the recursive relation in Eq.~\eqref{eq:jump_weights}, where the dependence on the jumping times $t_1,\dots,t_n$ is implicit.
    
    A similar procedure could in principle also be followed for the $\hat{\sigma}_z=\ket{\Phi_1}\bra{\Phi_1}-\ket{\Phi_0}\bra{\Phi_0}$ jump operator, leading to
    \begin{subequations}\label{eq:sz_weights}
    \begin{align}
        W_{O\mathrm{P}}^{(1)}(t;t_1) &= W_{O\mathrm{P}}(\bm{S}(t_1)) \left(1 + \sgn{(S_z(t_1)S'_z)}\right)|S'_z(t)| \nonumber\\
        &\quad- W_{O\mathrm{C}}(\bm{S}(t_1)) \left(S_x(t_1)S'_x + S_y(t_1)S'_y\right),\\
        W_{O\mathrm{C}}^{(1)}(t;t_1) &= W_{O\mathrm{P}}(\bm{S}(t_1)) \left(1 + \sgn{(S_z(t_1)S'_z)}\right) \nonumber\\
        &\quad- W_{O\mathrm{C}}(\bm{S}(t_1))\frac{3}{2} \left(S_x(t_1)S'_x + S_y(t_1)S'_y\right).
    \end{align}
    \end{subequations}
    However, as explained in the main text, we do not actually use this approach in practice. Appendix~\ref{ap:proof_eom} shows that simply rotating the spin vector around the $z$-axis without any reweighting also recovers the correct dynamics. Although both approaches [sampling $\bm{S}'$ from the Bloch sphere and reweighting using Eq.~\eqref{eq:sz_weights}, or rotating without reweighting] reproduce the correct reduced density-matrix evolution, reweighting can introduce negative weights and increase the number of trajectories necessary to obtain converged results. For this reason, we prefer the more efficient procedure of rotating the vector after $\hat{\sigma}_z$ jumps.

\subsection{Equations of motion}\label{ap:proof_eom}
    Equipped with the weighting factors, we now consider the evolution of the density matrix along a predetermined nuclear path.  
    We would like to recover the dynamics of Redfield theory [Eq.~\eqref{eq:redfield_secular_me_time}], which is
    \begin{subequations}
    \begin{align}
        \dot{\rho}_x(t) =& -\omega_{\rm LS}(t)\rho_y(t) + 2\tau(t)\rho_z(t) \nonumber\\
        &- \left[\frac{\gamma_+(t) + \gamma_-(t)}{2} + 2\gamma_z(t)\right] \rho_x(t), \label{eq:rhox_dot} \\
        \dot{\rho}_y(t) =& \omega_{\rm LS}(t)\rho_x(t) - \left[\frac{\gamma_+(t) + \gamma_-(t)}{2} + 2\gamma_z(t)\right] \rho_y(t), \\
        \dot{\rho}_z(t) =& -2\tau(t)\rho_x(t) - [\gamma_-(t) - \gamma_+(t)] \nonumber \\
        &- [\gamma_+(t) + \gamma_-(t)] \rho_z(t). \label{eq:rhoz_dot}
    \end{align} \label{eq:rho_dot}
    \end{subequations}
    
    For the equivalent stochastic LME expressed via the MASH kernel, using $\hat{\rho}(0) = \hat{\rho}_\mathrm{P} + \hat{\rho}_\mathrm{C}$, where $\hat{\rho}_\mathrm{P} = \frac{1}{2}(\hat{I} + \hat{\sigma}_z\rho_z(0))$ and $\hat{\rho}_\mathrm{C} = \frac{1}{2}(\hat{\sigma}_x\rho_x(0) + \hat{\sigma}_y\rho_y(0))$,
    \begin{align}
        \rho^{\rm MASH}_z(t) &= \int\!\!\cdots\!\!\int \dd \bm{S}^{(0)}\!\cdots \dd \bm{S}^{(n)} \, \rho_\mathrm{P}(\bm{S}^{(0)}) W^{(n)}_{\mathrm{P}\mathrm{P}}(t)\sgn{S^{(n)}_z(t)} \nonumber \\
        &+ \int\!\!\cdots\!\!\int \dd \bm{S}^{(0)}\!\cdots \dd \bm{S}^{(n)} \, \rho_\mathrm{C}(\bm{S}^{(0)}) W^{(n)}_\mathrm{CP}(t)\sgn{S^{(n)}_z(t)},\label{eq:rho_z_mash}
    \end{align}
    where the integrals are over the initial 
    spin vectors $\bm{S}^{(0)}$ and all of the sampled spin vectors after stochastic jumps. We also implicitly average over the number of stochastic jumps. $W^{(n)}_{\mu\nu}$ are the weighting factors after $n$ jumps, and $\bm{S}^{(n)}$ is the spin vector sampled after the $n$th jump.
    
    In Appendix~\ref{ap:proof_weight}, we have already shown how to modify the weighting factors to obtain the correct density matrix after stochastic jumps. Therefore, here we only need to consider the situation before any stochastic jumps:
    \begin{align}
        \rho^{\rm MASH}_z(t) &= \int \dd \bm{S} \, 2\rho_\mathrm{P}(\bm{S})|S_z(t)|\sgn{S_z(t)} \nonumber \\
        &+ \int \dd \bm{S} \, 3\rho_\mathrm{C}(\bm{S}) |S_z(t)|\sgn{S_z(t)}.
    \end{align}
    We can then obtain $\rho^{\rm MASH}_z(t+\dd t)$ and $\dot{\rho}^{\rm MASH}_z(t)$ using Eq.~\eqref{eq:evol_dist} and $|S_z|\sgn S_z = S_z$:
    \begin{align}
        \rho^{\rm MASH}_z(t+\dd t) &= \int \dd \bm{S} \, [2\rho_\mathrm{P}(\bm{S}) + 3\rho_\mathrm{C}(\bm{S})]  \nonumber \\
        &\quad\times\biggl[(1 - \Sigma(t)\dd t)(S_z(t) + \dot{S}_z(t) \dd t) \nonumber \\
        &\quad + \mathcal{P}_z(t)S_z(t) \dd t \nonumber \\
        &\quad + \mathcal{P}_-(t)|S_z(t)|\dd t \int \dd \bm{S}' 2|S'_z|h(-S'_z) \sgn{(S'_z)} \nonumber \\
        &\quad + \mathcal{P}_+(t)|S_z(t)|\dd t \int \dd \bm{S}' 2|S'_z|h(S'_z) \sgn{(S'_z)}\biggr].
    \end{align}
    The integrals over $\bm{S}'$ can be evaluated as 
    \begin{equation}
        \int \dd \bm{S}' \, 2|S'_z|h(\pm S'_z) \sgn{(S'_z)} = \pm 1,
    \end{equation}
    and combined with the terms coming from $S_z(t)\Sigma(t)\dd t$ as
    \begin{equation}
        \int \dd\bm{S} \, \mathcal{P}_\pm(t)\big[\pm|S_z(t)| - S_z(t)\big] = \pm 2 \int \dd\bm{S} \, \mathcal{P}_\pm(t)|S_z(t)|,
    \end{equation}
    where we used $\mathcal{P}_\pm(t)\propto h(\mp S_z(t))$ derived at the end of Appendix~\ref{ap:proof_weight}.

    Using $\rho^{\rm MASH}_z(t+\dd t) = \rho^{\rm MASH}_z(t) + \dot{\rho}^{\rm MASH}_z(t)\dd t + \mathcal{O}(\dd t^2)$, $\dot{S}_z$ from Eq.~\eqref{eq:spin_evol_sz} and the definition of $\rho_x^{\rm MASH}(t)$ given in Eq.~\eqref{eq:rho_x_mash} we get 
    \begin{align}
        \dot{\rho}^{\rm MASH}_z(t) &=-2\tau(t)\rho_x^{\rm MASH}(t) \nonumber \\
        &+ \int \dd \bm{S} \, \big[2\rho_\mathrm{P}(\bm{S}) + 3\rho_\mathrm{C}(\bm{S})\big] \times \big[\mathcal{G}_z(\bm{S}(t)) \nonumber \\
        &- 2\mathcal{P}_-(t)|S_z(t)| + 2\mathcal{P}_+(t)|S_z(t)|\big].
    \end{align}
    We therefore define $\mathcal{P}_\pm(t) = \gamma_\pm(t)h(\mp S_z(t))$ and $\mathcal{G}_z(\bm{S}) = 0$ such that we recover the correct behavior [Eq.~\eqref{eq:rhoz_dot}]:
    \begin{align}
        \dot{\rho}^{\rm MASH}_z(t) =& -2\tau(t)\rho^{\rm MASH}_x(t) \nonumber \\
        &-  [\gamma_-(t) - \gamma_+(t)] - [\gamma_+(t) + \gamma_-(t)]\rho^{\rm MASH}_z(t),
    \end{align}
    where we used $h(\pm S_z) = \tfrac{1}{2}(1 \pm\sgn S_z)$.
    
    Similarly for $\rho_x(t)$, the MASH approximation is
    \begin{align}
        \rho^{\rm MASH}_x(t) &= \int \dd \bm{S} \,2\rho_\mathrm{P}(\bm{S})S_x(t) + \int \dd \bm{S} \, 3\rho_\mathrm{C}(\bm{S}) S_x(t). \label{eq:rho_x_mash}
    \end{align}
    From this, we can derive $\rho^{\rm MASH}_x(t+\dd t)$ and subsequently $\dot{\rho}^{\rm MASH}_x(t)$ in the same way as we did for the $z$-component:
    \begin{align}
        \rho^{\rm MASH}_x(t+\dd t) &= \int \dd \bm{S} \, [2\rho_\mathrm{P}(\bm{S}) + 3\rho_\mathrm{C}(\bm{S})] \nonumber \\
        &\quad\times\biggl[(1 - \Sigma(t)\dd t)(S_x + \dot{S}_x dt)  - \mathcal{P}_z(t)S_x \dd t  \nonumber \\
        &\quad + \mathcal{P}_-(t)\dd t \int \dd \bm{S}' \, 2 S'_x \, h(-S'_z) \nonumber \\
        &\quad + \mathcal{P}_+(t)\dd t \int \dd \bm{S}' \,2S'_x \, h(S'_z)\biggr].
    \end{align}
    The integrals over $\bm{S}'$ vanish. Using $\mathcal{P}_\pm(t) = \gamma_\pm(t)h(\mp S_z(t))$, we obtain the following expression for the time derivative:
    \begin{align}
        \dot{\rho}^{\rm MASH}_x(t) &= -\omega_{\rm LS}(t) \rho_y^{\rm MASH}(t) + 2\tau(t)\rho_z^{\rm MASH}(t)\nonumber \\
        &\quad+ \int\dd \bm{S} \, \big[2\rho_\mathrm{P}(\bm{S}) + 3\rho_\mathrm{C}(\bm{S})\big]\times \big[\mathcal{G}_x(\bm{S}(t)) \nonumber \\
        &\quad - \gamma_-(t)h(S_z(t))S_x(t) - \gamma_+(t)h(-S_z(t))S_x(t)  \nonumber \\
        &\quad- 2\mathcal{P}_z(t) S_x(t)\big].
    \end{align}
    By choosing the probability $\mathcal{P}_z(t) = \gamma_z(t)$ and dissipative term $\mathcal{G}_x(\bm{S}(t)) = -\frac{1}{2}[\gamma_-(t) - \gamma_+(t)]S_x(t)\sgn S_z(t)$, we recover the correct behavior [Eq.~\eqref{eq:rhox_dot}]:
    \begin{align}
        \dot{\rho}^{\rm MASH}_x(t) &= -\omega_{\rm LS}(t) \rho^{\rm MASH}_y(t) + 2\tau(t)\rho^{\rm MASH}_z(t) \nonumber \\
        &- \left[\frac{\gamma_+(t) + \gamma_-(t)}{2} + 2\gamma_z(t)\right]\rho^{\rm MASH}_x(t).
    \end{align}
    A similar procedure for $\rho_y(t)$ allows us to determine $\mathcal{G}_y(\bm{S}(t)) = -\frac{1}{2}[\gamma_-(t) - \gamma_+(t)]S_y(t)\sgn S_z(t)$.
    
    These equations prove that the spin evolution presented in Eq.~\eqref{eq:spin_evol} does indeed reproduce the correct reduced density-matrix dynamics on an ensemble level along a given nuclear trajectory (or equivalently for a time-dependent Hamiltonian). 
    Therefore instead of the conventional stochastic unravelling of Sec.~\ref{sec:stochastic}, one can recover the same results as Redfield theory [Eq.~\eqref{eq:redfield_secular_me_time}] using an ensemble of spin vectors with weighting factors and correlation functions defined within the MASH framework.
    This provides the theoretical foundation on which the hybrid Redfield--MASH method is constructed.

    \section{Connection to QCLE-CME}\label{ap:qcle-cme}
    In this Appendix, we explain the connection between the hybrid Redfield--MASH method 
    and the ``QCLE inside a classical master equation,'' introduced in Refs.~\onlinecite{QCLE-CME-friction,QCLE-CME}.
    In our case, we take a partial Wigner transform of the secular Redfield equation [Eq.~\eqref{eq:redfield_secular_me_time}] to obtain the QCLE-CME time evolution of a general operator $\hat{\mathcal{B}}$ as
    \begin{align}
        \dot{\hat{\mathcal{B}}}(\bm{q},\bm{p},t) &= \ii \left[\Hs(t) + \Hls(t)  + \tau(t)\hat{\sigma}_y,\,\hat{\mathcal{B}}(\bm{q},\bm{p},t)\right] \nonumber \\
        &\quad+ \sum_{\nu} \gamma_\nu(t)\big(\hat{\sigma}_\nu\hat{\mathcal{B}}(\bm{q},\bm{p},t) \hat{\sigma}^\dagger_\nu - \tfrac{1}{2}\left[\hat{\sigma}^\dagger_\nu \hat{\sigma}_\nu, \hat{\mathcal{B}}(\bm{q},\bm{p},t)\right]_+\big) \nonumber \\
        &\quad+ \frac{1}{2}\sum_j \left[\hat{F}_j(\bm{q}),\frac{\partial \hat{\mathcal{B}}(\bm{q},\bm{p},t)}{\partial p_j}\right]_+ + \sum_j p_j \frac{\partial \hat{\mathcal{B}}(\bm{q},\bm{p},t)}{\partial q_j},\label{eq:qcle-cme}
    \end{align}
    where the quantum force operator is defined as 
    \begin{equation}
        \hat{F}_j(\bm{q}) = -\pder{\bar{V}_{\rm LS}(\bm{q})}{q_j}\hat{I} -\frac{1}{2}\pder{\omega_{\rm LS}(\bm{q})}{q_j}\hat{\sigma}_z + \omega_\mathrm{LS} (\bm{q})d_j(\bm{q})\hat{\sigma}_x.
    \end{equation}

    The first two terms on the right-hand side of Eq.~\eqref{eq:qcle-cme} describe the bare electronic dynamics, and are accurately described in the hybrid method thanks to the derivation in Appendix~\ref{ap:proof}. 
    The final two terms of Eq.~\eqref{eq:qcle-cme} are equivalent to those which appear in the standard QCLE, for which standard MASH dynamics is known to capture to first-order in time.\cite{MASH,MASHreview}
    In particular, using the classical force in the hybrid method [Eq.~\eqref{eq:mash_force}], it can be shown that one correctly captures third term of Eq.~\eqref{eq:qcle-cme} if Eq.~\eqref{eq:qcle_requirement} is obeyed, and we show in Appendix \ref{ap:QCLE} that this holds for almost all operator pairings using the new weighting factors. The fourth term is also exactly reproduced in the dynamics of the hybrid method using Eq.~\eqref{eq:mash_qdot}.
    In this way, we demonstrate that the hybrid Redfield--MASH method is equivalent to Eq.~\eqref{eq:qcle-cme} in the short-time limit except for the special case of coherence--momentum correlation functions.

\bibliography{references, references_new}

@String { jcp      = {J.~Chem. Phys.} }

@String { jctc     = {J.~Chem. Theory Comput.} }

@String { jpca     = {J.~Phys. Chem.~A} }

@String { jpcb     = {J.~Phys. Chem.~B} }

@String { jpcl     = {J.~Phys. Chem. Lett.} }

@String { prl      = {Phys. Rev. Lett.} }

@Article{thermalization,
  Title                    = {Detailed balance in mixed quantum--classical mapping approaches},
  Author                   = {Graziano Amati and Jonathan R Mannouch and Jeremy O Richardson},
  Journal                  = jcp,
  Year                     = {2023},
  Pages                    = {214114},
  Volume                   = {159},

  Arxivprefix              = {arXiv},
  Doi                      = {10.1063/5.0176291},
  Eprint                   = {2309.04686},
  Eprintclass              = {quant-ph}
}

@Misc{pyrho,
  Title                    = {{pyrho: A python package for reduced density matrix techniques}},

  Author                   = {Berkelbach,Timothy C.},
  HowPublished             = {\url{https://github.com/berkelbach-group/pyrho}}
}

@Book{OpenQuantum,
  Title                    = {The Theory of Open Quantum Systems},
  Author                   = {H.-P. Breuer and F. Petruccione},
  Publisher                = {Oxford University Press},
  Year                     = {2002},

  Address                  = {Oxford}
}

@Article{MASHcoh,
  Title                    = {{Simulating electronic coherences induced by conical intersections using MASH: Application to attosecond \mbox{X-ray} spectroscopy}},
  Author                   = {Daniele Furlanetto and Jeremy O Richardson},
  Journal                  = jpcl,
  Year                     = {2025},
  Pages                    = {6794-6800},
  Volume                   = {16},

  Arxivprefix              = {arXiv},
  Doi                      = {10.1021/acs.jpclett.5c01407},
  Eprint                   = {2505.05403},
  Eprintclass              = {physics.chem-ph}
}

@Article{MASHEOM,
  Title                    = {{Time-Reversible Implementation of MASH for Efficient Nonadiabatic Molecular Dynamics}},
  Author                   = {J Amira Geuther and Kasra Asnaashari and Jeremy O Richardson},
  Journal                  = jctc,
  Year                     = {2025},
  Pages                    = {2179-2188},
  Volume                   = {21},

  Archiveprefix            = {2412.15976},
  Doi                      = {10.1021/acs.jctc.4c01684}
}

@Article{Guo1996ZPE,
  Title                    = {Analysis of the zero-point energy problem in classical trajectory simulations},
  Author                   = {Guo, Yin and Thompson, Donald L and Sewell, Thomas D},
  Journal                  = {J.~Chem. Phys.},
  Year                     = {1996},
  Number                   = {2},
  Pages                    = {576--582},
  Volume                   = {104},
  Publisher                = {American Institute of Physics}
}

@Article{Habershon2009water,
  Title                    = {Zero point energy leakage in condensed phase dynamics: An assessment of quantum simulation methods for liquid water},
  Author                   = {Habershon, Scott and Manolopoulos, David E},
  Journal                  = {J.~Chem. Phys.},
  Year                     = {2009},
  Number                   = {24},
  Pages                    = {244518},
  Volume                   = {131},

  Doi                      = {10.1063/1.3276109},
  Publisher                = {American Institute of Physics}
}

@Article{HeadGordon1995friction,
  Title                    = {Molecular dynamics with electronic frictions},
  Author                   = {Head-Gordon, Martin and Tully, John C},
  Journal                  = {J.~Chem. Phys.},
  Year                     = {1995},
  Number                   = {23},
  Pages                    = {10137--10145},
  Volume                   = {103},
  Publisher                = {AIP}
}

@Article{Kapral2015QCL,
  Title                    = {Quantum dynamics in open quantum-classical systems},
  Author                   = {Kapral, Raymond},
  Journal                  = {J.~Phys.: Condens. Mat.},
  Year                     = {2015},
  Number                   = {7},
  Pages                    = {073201},
  Volume                   = {27},

  Doi                      = {10.1088/0953-8984/27/7/073201},
  Publisher                = {IOP Publishing}
}

@Article{HEOMclassical,
  Title                    = {Quantum nature of reactivity modification in vibrational polariton chemistry},
  Author                   = {Yaling Ke and Jeremy O Richardson},
  Journal                  = jcp,
  Year                     = {2024},
  Pages                    = {054104},
  Volume                   = {161},

  Doi                      = {10.1063/5.0220908}
}

@Article{cyclobutanone,
  Title                    = {{A MASH simulation of the photoexcited dynamics of cyclobutanone}},
  Author                   = {Joseph E. Lawrence and Imaad M. Ansari and Jonathan R. Mannouch and Meghna A. Manae and Kasra Asnaashari and Aaron Kelly and Jeremy O. Richardson},
  Journal                  = jcp,
  Year                     = {2024},
  Number                   = {174306},
  Pages                    = {174306},
  Volume                   = {160},

  Archiveprefix            = {arXiv},
  Doi                      = {10.1063/5.0203695},
  Eprint                   = {2402.10410},
  Eprintclass              = {physics.chem-ph}
}

@Article{MASHrates,
  Title                    = {{Recovering Marcus Theory Rates and Beyond without the Need for Decoherence Corrections: The Mapping Approach to Surface Hopping}},
  Author                   = {Joseph E Lawrence and Jonathan R Mannouch and Jeremy O Richardson},
  Journal                  = jpcl,
  Year                     = {2024},
  Pages                    = {707-716},
  Volume                   = {15},

  Arxivprefix              = {arXiv},
  Doi                      = {10.1021/acs.jpclett.3c03197},
  Eprint                   = {2311.08802},
  Eprintclass              = {physics.chem-ph}
}

@Article{unSMASH,
  Title                    = {A Size-Consistent Multi-State Mapping Approach to Surface Hopping},
  Author                   = {Joseph E Lawrence and Jonathan R Mannouch and Jeremy O Richardson},
  Journal                  = jcp,
  Year                     = {2024},
  Pages                    = {244112},
  Volume                   = {160},

  Doi                      = {10.1063/5.0208575}
}

@Article{Li2012fermions,
  Title                    = {{A Cartesian classical second-quantized many-electron Hamiltonian, for use with the semiclassical initial value representation}},
  Author                   = {Li, Bin and Miller, William H},
  Journal                  = {J.~Chem. Phys.},
  Year                     = {2012},
  Number                   = {15},
  Pages                    = {154107},
  Volume                   = {137},
  Publisher                = {AIP}
}

@Book{MahlerBook,
  Title                    = {Quantum Networks: Dynamics of Open Nanostructures},
  Author                   = {G. Mahler and V. A. Weberru\ss},
  Publisher                = {Springer},
  Year                     = {1995}
}

@Article{Mannouch2024coherence,
  Title                    = {Toward a Correct Description of Initial Electronic Coherence in Nonadiabatic Dynamics Simulations},
  Author                   = {Mannouch, Jonathan R and Kelly, Aaron},
  Journal                  = jpcl,
  Year                     = {2024},
  Pages                    = {11687--11695},
  Volume                   = {15},
  Publisher                = {ACS Publications}
}

@Article{MASH,
  Title                    = {A mapping approach to surface hopping},
  Author                   = {Jonathan R Mannouch and Jeremy O Richardson},
  Journal                  = jcp,
  Year                     = {2023},
  Pages                    = {104111},
  Volume                   = {158},

  Archiveprefix            = {arXiv},
  Doi                      = {10.1063/5.0139734},
  Eprint                   = {2212.11773},
  Primaryclass             = {physics.chem-ph}
}

@Article{Meyer1979nonadiabatic,
  Title                    = {A classical analog for electronic degrees of freedom in nonadiabatic collision processes},
  Author                   = {Meyer, H.-D. and Miller, W. H.},
  Journal                  = {J.~Chem. Phys.},
  Year                     = {1979},
  Number                   = {7},
  Pages                    = {3214--3223},
  Volume                   = {70},

  Doi                      = {10.1063/1.437910}
}

@Article{Miller2016Faraday,
  Title                    = {Classical molecular dynamics simulation of electronically non-adiabatic processes},
  Author                   = {Miller, William H and Cotton, Stephen J},
  Journal                  = {Faraday Discuss.},
  Year                     = {2016},
  Pages                    = {9--30},
  Volume                   = {195},

  Doi                      = {10.1039/C6FD00181E},
  Publisher                = {The Royal Society of Chemistry}
}

@Article{Miller1986fermions,
  Title                    = {{Classical models for electronic degrees of freedom: The second-quantized many-electron Hamiltonian}},
  Author                   = {Miller, William H and White, Kim A},
  Journal                  = {J.~Chem. Phys.},
  Year                     = {1986},
  Number                   = {9},
  Pages                    = {5059--5066},
  Volume                   = {84},
  Publisher                = {AIP}
}

@Article{Montoya2018fermions,
  Title                    = {On the exact continuous mapping of fermions},
  Author                   = {Montoya-Castillo, Andr{\'e}s and Markland, Thomas E},
  Journal                  = {Scientific reports},
  Year                     = {2018},
  Number                   = {1},
  Pages                    = {12929},
  Volume                   = {8},
  Publisher                = {Nature Publishing Group}
}

@Article{MASHreview,
  Title                    = {{Mapping Approach to Surface Hopping (MASH)}},
  Author                   = {Jeremy O. Richardson and Joseph E. Lawrence and Jonathan R. Mannouch},
  Journal                  = {Annu. Rev. Phys. Chem.},
  Year                     = {2025},
  Pages                    = {29},
  Volume                   = {76},

  Doi                      = {10.1146/annurev-physchem-082423-120631}
}

@Article{Runeson2024MASH,
  Title                    = {{Exciton dynamics from the mapping approach to surface hopping: Comparison with F{\"o}rster and Redfield theories}},
  Author                   = {Runeson, Johan E and Fay, Thomas P and Manolopoulos, David E},
  Journal                  = {Phys. Chem. Chem. Phys.},
  Year                     = {2024},
  Pages                    = {4929-4938},
  Volume                   = {26},

  Doi                      = {10.1039/D3CP05926J}
}

@Article{Runeson2023MASH,
  Title                    = {A multi-state mapping approach to surface hopping},
  Author                   = {Runeson, Johan E and Manolopoulos, David E},
  Journal                  = jcp,
  Year                     = {2023},
  Pages                    = {094115},
  Volume                   = {159},

  Doi                      = {10.1063/5.0158147}
}

@Article{multispin,
  Title                    = {Generalized spin mapping for quantum-classical dynamics},
  Author                   = {Johan E. Runeson and Jeremy O Richardson},
  Journal                  = {J. Chem. Phys.},
  Year                     = {2020},
  Pages                    = {084110},
  Volume                   = {152},

  Archiveprefix            = {arXiv},
  Doi                      = {10.1063/1.5143412},
  Eprint                   = {1912.10906},
  Primaryclass             = {physics.chem-ph}
}

@Article{spinmap,
  Title                    = {Spin-mapping approach for nonadiabatic molecular dynamics},
  Author                   = {Johan E. Runeson and Jeremy O. Richardson},
  Journal                  = {J. Chem. Phys.},
  Year                     = {2019},
  Pages                    = {044119},
  Volume                   = {151},

  Archiveprefix            = {arXiv},
  Doi                      = {10.1063/1.5100506},
  Eprint                   = {1904.08293},
  Primaryclass             = {physics.chem-ph},
  Shortauthor              = {Johan E. Runeson and Jeremy O. Richardson\corr}
}

@Article{identity,
  Title                    = {On the identity of the identity operator in nonadiabatic linearized semiclassical dynamics},
  Author                   = {Saller, Maximilian A C and Kelly, Aaron and Richardson, Jeremy O},
  Journal                  = {J. Chem. Phys.},
  Year                     = {2019},
  Pages                    = {071101},
  Volume                   = {150},

  Archiveprefix            = {arXiv},
  Doi                      = {10.1063/1.5082596},
  Eprint                   = {1811.08830},
  Primaryclass             = {physics.chem-ph},
  Shortauthor              = {Saller, Maximilian A C and Kelly, Aaron and Richardson\corr, Jeremy O}
}

@Article{Siders1981inverted,
  Title                    = {Quantum effects for electron-transfer reactions in the ``inverted region''},
  Author                   = {Siders, P and Marcus, R A},
  Journal                  = {J.~Am. Chem. Soc.},
  Year                     = {1981},
  Number                   = {4},
  Pages                    = {748--752},
  Volume                   = {103},

  Doi                      = {10.1021/ja00394a004},
  Publisher                = {ACS Publications}
}

@Article{Siders1981quantum,
  Title                    = {Quantum effects in electron-transfer reactions},
  Author                   = {Siders, P and Marcus, R A},
  Journal                  = {J.~Am. Chem. Soc.},
  Year                     = {1981},
  Number                   = {4},
  Pages                    = {741--747},
  Volume                   = {103},

  Doi                      = {10.1021/ja00394a003},
  Publisher                = {ACS Publications}
}

@Article{Stock1997mapping,
  Title                    = {Semiclassical description of nonadiabatic quantum dynamics},
  Author                   = {Stock, G. and Thoss, M.},
  Journal                  = prl,
  Year                     = {1997},
  Number                   = {4},
  Pages                    = {578--581},
  Volume                   = {78},

  Doi                      = {10.1103/PhysRevLett.78.578},
  Publisher                = {APS}
}

@Article{Tanimura2020HEOM,
  Title                    = {{Numerically ``exact'' approach to open quantum dynamics: The hierarchical equations of motion (HEOM)}},
  Author                   = {Tanimura, Yoshitaka},
  Journal                  = {J. Chem. Phys.},
  Year                     = {2020},
  Number                   = {2},
  Pages                    = {020901},
  Volume                   = {153},

  Doi                      = {10.1063/5.0011599},
  Publisher                = {AIP Publishing LLC}
}

@Article{Tully1990hopping,
  Title                    = {Molecular dynamics with electronic transitions},
  Author                   = {Tully, J. C.},
  Journal                  = {J.~Chem. Phys.},
  Year                     = {1990},
  Pages                    = {1061--1071},
  Volume                   = {93},

  Doi                      = {10.1063/1.459170}
}

@Book{Zwanzig,
  Title                    = {Nonequilibrium Statistical Mechanics},
  Author                   = {Robert Zwanzig},
  Publisher                = {Oxford University Press},
  Year                     = {2001},

  Address                  = {New York}
}

@Article{Runeson2024semiconductors,
  author    = {Runeson, Johan E and Drayton, Thomas JG and Manolopoulos, David E},
  journal   = {J. Chem. Phys.},
  title     = {Charge transport in organic semiconductors from the mapping approach to surface hopping},
  volume    = {161},
  year      = {2024},
  doi       = {10.1063/5.0226001},
  number    = {14},
  publisher = {AIP Publishing},
}

@Article{Mannouch2024MASH,
  Title                    = {{Quantum quality with classical cost: Ab initio nonadiabatic dynamics simulations using the mapping approach to surface hopping}},
  Author                   = {Mannouch, Jonathan R and Kelly, Aaron},
  Journal                  = {J. Phys. Chem. Lett.},
  Year                     = {2024},
  Pages                    = {5814--5823},
  Volume                   = {15},

  doi                      = {10.1021/acs.jpclett.4c00535},
  Publisher                = {ACS Publications}
}

@Article{Liu2016mapping,
  author    = {Liu, Jian},
  journal   = {J. Chem. Phys.},
  title     = {A unified theoretical framework for mapping models for the multi-state Hamiltonian},
  volume    = {145},
  year      = {2016},
  number    = {20},
  publisher = {AIP Publishing},
}

@InCollection{Redfield1965relaxation,
  author    = {Redfield, Alfred G},
  booktitle = {Adv. Magn. Opt. Reson.},
  title     = {The theory of relaxation processes},
  pages     = {1--32},
  publisher = {Elsevier},
  volume    = {1},
  year      = {1965},
}

@mastersthesis{mash-thesis,
      title={An Analysis of the Mapping Approach to Surface Hopping}, 
      author={Jan Vavřín},
      year={2025},
      eprint={2507.02970},
      archivePrefix={arXiv},
      primaryClass={physics.chem-ph},
      url={https://arxiv.org/abs/2507.02970}, 
school = {University of Cambridge},
}

@book{Carmichael1,
  title={Statistical methods in quantum optics 1: Master Equations and Fokker-Planck Equations},
  author={Carmichael, Howard J},
  year={1999},
  publisher={Springer}
}

@book{Carmichael2,
  title={Statistical methods in quantum optics 2: Non-classical fields},
  author={Carmichael, Howard J},
  year={2008},
  publisher={Springer}
}

@article{RSH,
    author = {Pérez-Escribano, Manuel and Jankowska, Joanna and Granucci, Giovanni and Escudero, Daniel},
    title = {{The radiative surface hopping (RSH) algorithm: Capturing fluorescence events in molecular systems within a semi-classical non-adiabatic molecular dynamics framework}},
    journal = jcp,
    volume = {158},
    number = {12},
    pages = {124104},
    year = {2023},
    month = {03},
    issn = {0021-9606},
    doi = {10.1063/5.0139516}
}

@article{IESH,
    author = {Shenvi, Neil and Roy, Sharani and Tully, John C.},
    title = {Nonadiabatic dynamics at metal surfaces: Independent-electron surface hopping},
    journal = jcp,
    volume = {130},
    number = {17},
    pages = {174107},
    year = {2009},
    month = {05},
    issn = {0021-9606},
    doi = {10.1063/1.3125436}
}

@article{IESH2,
    author = {Gardner, James and Corken, Daniel and Janke, Svenja M. and Habershon, Scott and Maurer, Reinhard J.},
    title = {Efficient implementation and performance analysis of the independent electron surface hopping method for dynamics at metal surfaces},
    journal = jcp,
    volume = {158},
    number = {6},
    pages = {064101},
    year = {2023},
    month = {02},
    issn = {0021-9606},
    doi = {10.1063/5.0137137}
}

@article{IESH3,
author = {De, Priyam Kumar and Jain, Amber},
title = {Metal-Induced Fast Vibrational Energy Relaxation: Quantum Nuclear Effects Captured in Diabatic Independent Electron Surface Hopping {(IESH-D)} Method},
journal = jpca,
volume = {127},
number = {18},
pages = {4166-4179},
year = {2023},
doi = {10.1021/acs.jpca.3c00054}
}

@article{QCLE-CME,
author = {Dou, Wenjie and Subotnik, Joseph E.},
title = {A Generalized Surface Hopping Algorithm To Model Nonadiabatic Dynamics near Metal Surfaces: The Case of Multiple Electronic Orbitals},
journal = jctc,
volume = {13},
number = {6},
pages = {2430-2439},
year = {2017},
doi = {10.1021/acs.jctc.7b00094}
}

@article{LME-SH,
author = {Wang, Yi-Siang and Nijjar, Parmeet and Zhou, Xin and Bondar, Denys I. and Prezhdo, Oleg V.},
title = {{Combining Lindblad Master Equation and Surface Hopping to Evolve Distributions of Quantum Particles}},
journal = jpcb,
volume = {124},
number = {21},
pages = {4326-4337},
year = {2020},
doi = {10.1021/acs.jpcb.0c03030},
}

@article{FR-SH,
author = {Wang, Yu and Mosallanejad, Vahid and Liu, Wei and Dou, Wenjie},
title = {{Nonadiabatic Dynamics near Metal Surfaces with Periodic Drivings: A Generalized Surface Hopping in Floquet Representation}},
journal = jctc,
volume = {20},
number = {2},
pages = {644-650},
year = {2024},
doi = {10.1021/acs.jctc.3c01263}
}

@article{SL-FSSH,
    author = {Ouyang, Wenjun and Dou, Wenjie and Subotnik, Joseph E.},
    title = {{Surface hopping with a manifold of electronic states. I. Incorporating surface-leaking to capture lifetimes}},
    journal = jcp,
    volume = {142},
    number = {8},
    pages = {084109},
    year = {2015},
    month = {02},
    issn = {0021-9606},
    doi = {10.1063/1.4908032}
}

@article{CME,
    author = {Dou, Wenjie and Nitzan, Abraham and Subotnik, Joseph E.},
    title = {{Surface hopping with a manifold of electronic states. II. Application to the many-body Anderson-Holstein model}},
    journal = jcp,
    volume = {142},
    number = {8},
    pages = {084110},
    year = {2015},
    month = {02},
    issn = {0021-9606},
    doi = {10.1063/1.4908034}
}

@article{CME2,
    author = {Dou, Wenjie and Nitzan, Abraham and Subotnik, Joseph E.},
    title = {{Surface hopping with a manifold of electronic states. III. Transients, broadening, and the Marcus picture}},
    journal = jcp,
    volume = {142},
    number = {23},
    pages = {234106},
    year = {2015},
    month = {06},
    issn = {0021-9606},
    doi = {10.1063/1.4922513}
}

@article{BerkelbachFrozen,
    author = {Fetherolf, Jonathan H. and Berkelbach, Timothy C.},
    title = {Linear and nonlinear spectroscopy from quantum master equations},
    journal = jcp,
    volume = {147},
    number = {24},
    pages = {244109},
    year = {2017},
    month = {12},
    issn = {0021-9606},
    doi = {10.1063/1.5006824}
}

@article{HEOMThoss,
    author = {Ke, Yaling and Erpenbeck, André and Peskin, Uri and Thoss, Michael},
    title = {Unraveling current-induced dissociation mechanisms in single-molecule junctions},
    journal = jcp,
    volume = {154},
    number = {23},
    pages = {234702},
    year = {2021},
    month = {06},
    issn = {0021-9606},
    doi = {10.1063/5.0053828}
}

@book{quantum-biology, 
    title={Quantum Effects in Biology}, 
    author={Y. Omar and M. B. Plenio},
    editor={M. Mohseni},
    address={Cambridge},
    publisher={Cambridge University Press},
    doi={https://doi.org/10.1017/CBO9780511863189},
    year={2014}
}

@book{GonzalezLindh,
    publisher = {John Wiley \& Sons, Ltd},
    title = {Quantum Chemistry and Dynamics of Excited States},
    editor = {Leticia González, Roland Lindh},
    doi = {https://doi.org/10.1002/9781119417774},
    year = {2020}
}

@article{sun2021bosonic,
  title={A bosonic perspective on the classical mapping of fermionic quantum dynamics},
  author={Sun, Jing and Sasmal, Sudip and Vendrell, Oriol},
  journal=jcp,
  volume={155},
  number={13},
  year={2021},
  pages={134110},
  doi={10.1063/5.0066740},
  publisher={AIP Publishing}
}

@article{spin-baths,
doi = {10.1088/0034-4885/63/4/204},
year = {2000},
month = {apr},
volume = {63},
number = {4},
pages = {669},
author = {N V Prokof'ev and P C E Stamp},
title = {Theory 
of the spin bath},
journal = {Rep.~Prog.~Phys.},
}

@book{quantum-optics, 
place={Cambridge}, 
title={Quantum Optics}, 
publisher={Cambridge University Press}, 
author={Scully, Marlan O. and Zubairy, M. Suhail}, 
year={1997}}

@article{nuclear-spin-bath,
  title = {Quantum theory for electron spin decoherence induced by nuclear spin dynamics in semiconductor quantum computer architectures: Spectral diffusion of localized electron spins in the nuclear solid-state environment},
  author = {Witzel, W. M. and Das Sarma, S.},
  journal = {Phys. Rev. B},
  volume = {74},
  issue = {3},
  pages = {035322},
  numpages = {24},
  year = {2006},
  month = {Jul},
  publisher = {American Physical Society},
  doi = {10.1103/PhysRevB.74.035322}
}

@article{botticelli2025semi,
  title={A semi-focused multi-state variant of the mapping approach to surface hopping},
  author={Botticelli, Samuele and Accomasso, Davide and Granucci, Giovanni},
  journal={ChemRxiv},
  doi={10.26434/chemrxiv-2025-9tg9g},
  year={2025}
}

@article{
Runeson2025NQE,
author = {Johan E. Runeson  and David E. Manolopoulos },
title = {Nuclear quantum effects slow down the energy transfer in biological light-harvesting complexes},
journal = {Sci. Adv.},
volume = {11},
number = {23},
pages = {eadw4798},
year = {2025},
doi = {10.1126/sciadv.adw4798}}

@article{lamb-shift-div,
  title = {The Electromagnetic Shift of Energy Levels},
  author = {Bethe, H. A.},
  journal = {Phys. Rev.},
  volume = {72},
  issue = {4},
  pages = {339--341},
  numpages = {0},
  year = {1947},
  month = {Aug},
  publisher = {American Physical Society},
  doi = {10.1103/PhysRev.72.339},
}

@Inbook{Rivas2012,
author="Rivas, {\`A}ngel
and Huelga, Susana F.",
title="Microscopic Description: Markovian Case",
bookTitle="Open Quantum Systems: An Introduction",
year="2012",
publisher="Springer Berlin Heidelberg",
address="Berlin, Heidelberg",
pages="49--80",
isbn="978-3-642-23354-8",
doi="10.1007/978-3-642-23354-8_5"
}

@Inbook{Kloeden1992,
author="Kloeden, Peter E.
and Platen, Eckhard",
title="Stochastic Differential Equations",
bookTitle="Numerical Solution of Stochastic Differential Equations",
year="1992",
publisher="Springer Berlin Heidelberg",
address="Berlin, Heidelberg",
isbn="978-3-662-12616-5",
doi="10.1007/978-3-662-12616-5_4"
}

@article{quantum-redfield,
    author = {Kühl, Axel and Domcke, Wolfgang},
    title = {{Multilevel Redfield description of the dissipative dynamics at conical intersections}},
    journal = jcp,
    volume = {116},
    number = {1},
    pages = {263-274},
    year = {2002},
    month = {01},
    issn = {0021-9606},
    doi = {10.1063/1.1423326}
}

@article{Warshel1976retinol,
  title={Bicycle-pedal model for the first step in the vision process},
  author={Warshel, Arieh},
  journal={Nature},
  volume={260},
  number={5553},
  pages={679--683},
  year={1976},
  publisher={Nature Publishing Group UK London}
}

@article{ZPE1,
author={Hsieh, Ming-Hsiu
and Krotz, Alex
and Tempelaar, Roel},
title={Focused Sampling for Low-Cost and Accurate Ehrenfest Modeling of Cavity Quantum Electrodynamics},
journal=jctc,
year={2025},
month={Nov},
day={28},
publisher={American Chemical Society}
}

@article{miller1989simple,
  title={A simple model for correcting the zero point energy problem in classical trajectory simulations of polyatomic molecules},
  author={Miller, William H and Hase, William L and Darling, Cynthia L},
  journal=jcp,
  volume={91},
  number={5},
  pages={2863--2868},
  year={1989},
  publisher={American Institute of Physics}
}

@article{bowman1989method,
  title={A method to constrain vibrational energy in quasiclassical trajectory calculations},
  author={Bowman, Joel M and Gazdy, Bela and Sun, Qiyan},
  journal=jcp,
  volume={91},
  number={5},
  year={1989},
  publisher={Department of Chemistry, Emory University, Atlanta, Georgia 30322 (US)}
}

@article{Varandas1994ZPE,
title = {A novel non-active model to account for the leak of zero-point energy in trajectory calculations. Application to {H + O$_2$} reaction near threshold},
journal = {Chem.~Phys.~Lett.},
volume = {225},
number = {1},
pages = {18-27},
year = {1994},
issn = {0009-2614},
doi = {https://doi.org/10.1016/0009-2614(94)00620-2},
author = {A.J.C. Varandas}
}

@article{BenNun1996ZPE,
    author = {Ben‐Nun, M. and Levine, R. D.},
    title = {On the zero point energy in classical trajectory computations},
    journal = jcp,
    volume = {105},
    number = {18},
    pages = {8136-8141},
    year = {1996},
    month = {11},
    issn = {0021-9606},
    doi = {10.1063/1.472668}
}

@article{QCLE-CME-friction,
    author = {Dou, Wenjie and Subotnik, Joseph E.},
    title = {A many-body states picture of electronic friction: The case of multiple orbitals and multiple electronic states},
    journal = jcp,
    volume = {145},
    number = {5},
    pages = {054102},
    year = {2016},
    month = {08},
    issn = {0021-9606},
    doi = {10.1063/1.4959604},
}

\end{document}